\documentclass[final,5p,times,twocolumn,authoryear]{elsarticle}

\usepackage[authoryear]{natbib}
\usepackage{graphicx}
\usepackage{subcaption}
\usepackage{amssymb,amsmath}
\usepackage{pdflscape}

\usepackage{lipsum}
\usepackage{epsfig}
\usepackage{epstopdf}
\usepackage[utf8]{inputenc}
\usepackage{newunicodechar}
\usepackage{textgreek}
\usepackage{physics}
\usepackage{inputenc}

\usepackage{afterpage}

\usepackage{amsthm}

\usepackage{multirow}
\usepackage{array}
\usepackage{url} 
\usepackage{xcolor}
\usepackage{adjustbox}
\usepackage{soul}
\usepackage{orcidlink}

\journal{High Energy Astrophysics}

\begin{document}

\begin{frontmatter}

%% Title, authors and addresses

%% use the tnoteref command within \title for footnotes;
%% use the tnotetext command for theassociated footnote;
%% use the fnref command within \author or \affiliation for footnotes;
%% use the fntext command for theassociated footnote;
%% use the corref command within \author for corresponding author footnotes;
%% use the cortext command for theassociated footnote;
%% use the ead command for the email address,
%% and the form \ead[url] for the home page:
%% \title{Title\tnoteref{label1}}
%% \tnotetext[label1]{}
%% \author{Name\corref{cor1}\fnref{label2}}
%% \ead{email address}
%% \ead[url]{home page}
%% \fntext[label2]{}
%% \cortext[cor1]{Zahir Shah}
%% \affiliation{organization={},
%%            addressline={}, 
%%            city={},
%%            postcode={}, 
%%            state={},
%%            country={}}
%% \fntext[label3]{}

%\title{Transition Blazar S5 1803+784: Multi-Wavelength Variability and Emission Mechanisms}
\title{Study of Multi-Wavelength Variability, Emission Mechanism and Quasi-Periodic Oscillation for Transition Blazar S5 1803+784}
%% use optional labels to link authors explicitly to addresses:

%\author[first]{Javaid Tantry}
%\affiliation[first]{{Department of Physics, University of Kashmir,}, city={Srinagar},
%           postcode={190006}, 
%            state={Kashmir},
%            country={India}}

%\author[second]{Zahir Shah}
%\affiliation[second]{{Department of Physics, Central University of Kashmir,},
%            addressline={Ganderbal}, 
%            city={Varanasi},
%           postcode={191131}, 
%            state={Kashmir},
%            country={India}}  

%\author[first]{Naseer Iqbal}

%%%%%%%%%%%%%%%%%%%%%%%%%%%%%%%%%%%%%%%%   Demo

\author[first]{Javaid Tantry\,\orcidlink{0009-0002-5150-3604}}
\ead{javaidtantray9@gmail.com}

\author[second]{Ajay Sharma\,\orcidlink{0000-0002-5221-0822}}
\ead{ajjjkhoj@gmail.com}

\author[third]{Zahir Shah\,\orcidlink{0000-0003-1458-4396}}
\ead{shahzahir4@gmail.com}

\author[first]{Naseer Iqbal\,\orcidlink{0000-0002-3100-8393}}
\ead{dni_phtr@kashmiruniversity.ac.in}

\author[third]{Debanjan Bose\,\orcidlink{0000-0003-1071-5854}}
\ead{debanjan.tifr@gmail.com}

\affiliation[first]{{Department of Physics,}, {University of Kashmir,}, 
            city={Srinagar},
            postcode={190006}, 
            state={Kashmir},
            country={India}}

\affiliation[second]{{S N Bose National Centre for Basic Sciences,},addressline={Block JD, Salt Lake}, city={Kolkata},
           postcode={700106}, 
            state={West Bengal},
            country={India}}

\affiliation[third]{{Department of Physics, Central University of Kashmir,}, 
            addressline={Ganderbal}, 
            postcode={191131}, 
            state={Kashmir},
            country={India}} 

\cortext[cor1]{Corresponding author: Javaid Tantry, Ajay Sharma, Debanjan Bose }

%%%%%%%%%%%%%%%%%%%%%%%%%%%%%%%%%%%%%%%%%%%%%

\begin{abstract}
This work present the results of a multi-epoch observational study of the blazar S5\,1803+784, carried out from 2019 to 2023. The analysis is based on simultaneous data obtained from the Swift/UVOT/XRT, ASAS-SN, and Fermi-LAT instruments. A historically high $\gamma$-ray flux  observed for this source on march 2022 ($\mathrm{2.26\pm0.062)\times10^{-6}~phcm^{-2}s^{-1}}$. This study investigates the $\gamma$-ray emission from a blazar, revealing a dynamic light curve with four distinct flux states: quiescent and high-flux by using the Bayesian Blocks (BB) algorithm. A potential transient quasi-periodic signal with an oscillation timescale of $\sim$411 days was identified, showing a local significance level surpassing 99.7$\%$ from the Lomb-Scargle Periodogram (LSP) and Damped Random Walk (DRW) analysis and exceeds 99.5$\%$ from the Weighted Wavelet Z-Transform (WWZ) analysis. The observed QPO was confirmed through an autoregressive process (AR(1)), with a significance level exceeding 99$\%$, suggesting a potential physical mechanism for such oscillations involves a helical motion of a magnetic plasma blob within the relativistic jet. Log parabola modeling of the $\gamma$-ray spectrum revealed a photon index ($\alpha_\gamma$) variation of 1.65$\pm$0.41 to 2.48$\pm$0.09 with a steepening slope, potentially indicative of particle cooling, changes in radiative processes, or modifications in the physical parameters. The $\alpha_\gamma$ of 2.48$\pm$0.09 may hint at an evolutionary transition state from BL\,Lac to FSRQ. A comparative analysis of variability across different energy bands reveals that Optical/UV and GeV emissions display greater variability compared to X-rays.  Broadband SED modeling shows that within a one-zone leptonic framework, the SSC model accurately reproduces flux states without external Compton contributions, highlighting magnetic fields crucial role.
\end{abstract} 
%%Graphical abstract
%\begin{graphicalabstract}
%\includegraphics{grabs}
%\end{graphicalabstract}

%%Research highlights
%\begin{highlights}
%\item Research highlight 1
%\item Research highlight 2
%\end{highlights}

\begin{keyword}
galaxies: active \sep galaxies:  BL Lacertae objects: S5\,1803+784 \sep jets \sep radiation mechanisms: non-thermal - gamma-rays \sep galaxies:  Jets; Active

%% keywords here, in the form: keyword \sep keyword, up to a maximum of 6 keywords
%% PACS codes here, in the form: \PACS code \sep code

%% MSC codes here, in the form: \MSC code \sep code
%% or \MSC[2008] code \sep code (2000 is the default)
\end{keyword}

\end{frontmatter}

%\tableofcontents

%% \linenumbers

%% main text

\section{Introduction}
\label{introduction}
Blazars are a peculiar class of radio-loud AGNs, distinguished by relativistic jets that are closely aligned with the observer's line of sight \cite{urry1995unified}. The relativistic jets in the blazar are efficient particle accelerators and emit non-thermal radiation across the electromagnetic spectrum. This non-thermal emission from the relativistic jets dominates the entire broadband spectra \cite{1995MNRAS.273..583D, urry1995unified}.
Strong $\gamma-$ray emissions, a high degree of polarization, and frequent variability in flux and spectra are some other characteristics of blazars \cite{wills1992, ackermann2015, Fan2018, 2020ApJ...892..105A, Zhang_2021}.  

The broadband spectral energy distribution (SED) of blazars shows a two-humped structure, with the low-energy component (peaking in the submillimeter to soft X-ray range) attributed to synchrotron radiation from relativistic electrons in the jet, while the high-energy component (peaking at MeV to TeV energies) is explained by using either the leptonic model or hadronic model \cite{abdo2010spectral}. 
%remains a topic of ongoing debate, with multiple theoretical explanations proposed \citep{2010ApJ...716...30A}.  Two prominent theoretical frameworks, leptonic and hadronic, are extensively utilized. 
According to the leptonic model, 
the high-energy component is attributed to the inverse Compton scattering of low-energy photons. This can involve low-energy synchrotron photons within the jet, known as the synchrotron self-Compton (SSC) process, or photons originating from external sources outside the jet (such as from the accretion disk, broad line region (BLR) and Dusty Torus), referred to as the external Compton process \cite{2013ApJ...763..134F, 1996ApJ...461..657B, 2002ApJ...575..667D, Dermer_2009}. Alternatively, hadronic models could also explain the high energy emission in blazars. The high energy peak may be the consequence of secondary decay products of charged particles, primarily pions, or synchrotron radiation from protons \cite{2013ApJ...768...54B}. 

%(LSP; $\nu_p <  10^{14}$ Hz), intermediate-synchrotron peaked (ISP; $10^{14} < \nu_p <  10^{15}$) and high synchrotron peaked (HSP; $\nu_p > 10^{15}$)
%$$$$$$$$$$$$$$$$$$$$$$$$$$$$$$$$$$$$$$$$$$$$
%\textcolor{blue}{Blazars are often classified as BL Lacertae objects (BL Lacs), and flat-spectrum radio quasars (FSRQs) based on their spectral properties. In their optical spectrum, FSRQs show strong emission line characteristics, while BL Lacs have weak or no emission line features \citep{urry1995unified}. Based on the peak frequency of the synchrotron component, blazars are also classified into low synchrotron peak ($\rm{LSP}; \nu_p < 10^{14} Hz$), intermediate-synchrotron peaked ($\rm{ISP}; 10^{14} < \nu_p < 10^{15} Hz$), and high synchrotron peaked ($\rm{HSP};  \nu_p > 10^{15} Hz$) blazars \citep{1995ApJ...444..567P}. The FSRQs belong to the LSP category, while a large fraction of  BL Lacs are HSP sources \citep{2020A&A...634A..80R}. Additionally, blazars that exhibit properties of both FSRQs and BL Lacs in varying flux states are known as transition blazars/changing look blazars (CLB) \citep{2013MNRAS.432L..66G, 2021ApJ...913..146M, Xiao_2022, 2024A&A...681A.116P, 2024A&A...685A.140R}. The shifting poses significant challenges to the AGN unification model \citep{2021ApJ...913..146M, Peña-Herazo_2021}. A promising explanation for this transition is a sudden change in accretion ratio.}
%$$$$$$$$$$$$$$$$$$$$$$$$$$$$$$$$$$$$$$$$$$$$ 

Blazars are often classified as BL Lacertae objects (BL Lacs), and flat-spectrum radio quasars (FSRQs) based on their spectral properties. In their optical spectrum, FSRQs show strong emission line characteristics, while BL Lacs have weak or no emission line features \citep{urry1995unified}. Based on the peak frequency of the synchrotron component, blazars are also classified into low synchrotron peak (LSP; $\nu_p <  10^{14}$ Hz), intermediate-synchrotron peaked (ISP; $10^{14} < \nu_p <  10^{15}$) and high synchrotron peaked (HSP; $\nu_p > 10^{15}$) blazars \citep{1995ApJ...444..567P}. The FSRQs belong to the LSP category, while a large fraction of  BL Lacs are HSP sources \citep{2020A&A...634A..80R}. Additionally, blazars that exhibit properties of both FSRQs and BL Lacs in varying flux states are known as transition blazars/changing look blazars (CLB) \citep{2013MNRAS.432L..66G, 2021ApJ...913..146M, Xiao_2022, 2024A&A...681A.116P, 2024A&A...685A.140R}. The shifting poses significant challenges to the AGN unification model \citep{2021ApJ...913..146M, Peña-Herazo_2021}. A promising explanation for this transition is a sudden change in accretion ratio.\par

Blazars exhibit stochastic flux variability across multiple wavelengths, with timescales ranging from minutes to years. However, some blazars exhibit persistent or transient quasi-periodic variability \cite{sobolewska2014stochastic, gierlinski2008periodicity, abdo2010gamma, king2013quasi, alston2014detection, ackermann2015multiwavelength, penil2022evidence, ren2023quasi, sharma2024detection}. The $\gamma$-ray emission in blazars is believed to originate from their relativistic jets, making the investigation of quasi-periodic signals in this band crucial for understanding jet physics and particle acceleration mechanisms. The continuous sky monitoring by Fermi's Large Area Telescope (LAT) provides an excellent opportunity to explore such features in $\gamma$-rays using long-term temporal observations. Notably, the blazar PG 1553+113 was the first $\gamma$-ray source reported to exhibit a quasi-periodic oscillation (QPO) with a period of 2.18$\pm$0.08 years, completing three full cycles \cite{ackermann2015multiwavelength}. It is believed that the blazer is in a binary system, thus a persistent QPO source \cite{tavani2018blazar}. Since its discovery, this source has become an ideal candidate for continuous monitoring in $\gamma$-rays and across other wavelengths. Numerous transient and persistent quasi-periodic oscillations (QPOs) have been reported in the $\gamma$-ray emission of blazars, with oscillation timescales ranging from days to years \cite{zhou201834, benkhali2020evaluating, sarkar2020multi, sarkar2021multiwaveband, penil2022evidence, das2023detection, banerjee2023detection, prince2023quasi, li2023radio, sharma2024detection}. \cite{penil2022evidence} analyzed a sample of 24 blazars using approximately 12 years of Fermi-LAT observations, identifying five sources with persistent periodic behavior at high significance levels. \cite{zhou201834} were the first to report a significant transient QPO in a blazar, with a period of $\sim$34.5 days and a significance exceeding 4.6$\sigma$. More recently, \cite{ren2023quasi} conducted a QPO study on a sample of 35 AGNs observed in $\gamma$-rays and identified a transient QPO with a period of $\sim$39 days in the source B2 1520+31, persisting for 17 complete cycles. Additionally, several studies have documented both transient and persistent multi-wavelength QPOs in blazars. A widely considered physical explanation for these $\gamma$-ray QPOs involves the helical jet model, where the periodic variation in the viewing angle of the emission region modulates the observed flux. However, the precise physical mechanisms driving this behavior remain elusive.\par

Blazar S5\,1803+78 is a BL\,Lac type identified by \cite{1981ApJ...247L..53B}. The redshift of this blazar is z = 0.684, determined by \citet{Lawrence_1996} and confirmed by  \cite{rector2001properties}
. The source has been identified as LSP in the AGN unification scheme (\cite{urry1995unified}). Initially, the source was extensively used for geodynamics studies \cite{1999NewAR..43..691G, gabuzda2000astrophysical}.
Later, optical and radio study of this source was carried out by \citet{Nesci_2002, 2012AcPol..52a..39N} during the optical active state of the source. During that period, the radio (8.4 GHz) light curves showed modest fluctuations lacking periodic signatures. Interestingly,  \citet{2018MNRAS.478..359K} investigated the relativistic jet dynamics using 30 VLBI observations, revealing a significant 6-year periodicity. However, previous long-term optical observations showed no evidence of periodicity \cite{Nesci_2002, Nesci_Maselli_Montagni_Sclavi_2012}. 
 Recently, \citet{BUTUZOVA202519} studied the variability evolution of this source using TESS observations, finding that its variability timescales span from several hours to days. The authors suggested that such a pattern of variability can be attributed to contributions from different regions within the emission zone. Alternatively, it may also be explained by the continuous formation and evolution of sub-components in the emitting region, each with varying sizes and Doppler factors.
%The authors suggested that short-term variability arises from differing Doppler factors across various volume elements within the optical emission region. 
Another study on this source examined morphological changes in its radio structure following the $\gamma$-ray flares, during the period between MJD 59063.5 -- 59120.5, and reported the emergence of two new components from the core propagated outwards \cite{10.1093/mnras/stab501}.
The broadband spectra of the source during intense flaring episodes were modeled using a single-zone leptonic framework that incorporated external Compton scattering from a dusty torus, providing constraints on the underlying physical parameters \cite{inproceedings}. Furthermore, a multi-wavelength study spanning MJD 58727 -- 59419, utilizing data from Fermi-LAT, SWIFT/XRT/UVOT, NuSTAR, and TUBITAK observations, explored the source's high-flux state. Although simultaneous optical and X-ray data were unavailable, the study revealed that blazar flares exhibit day-scale variability with similar rise and decay timescales, indicating a compact emission region near the central engine \cite{10.1093/mnras/stac1009}. Notably, a bright optical flare was observed between May 2020 and May 2021, however, the optical variability during this period was less pronounced compared to the non-flaring state \cite{Agarwal_2022}.\\
Our study presents a comprehensive analysis of the blazar S5\,1803+784, utilizing simultaneous observations from Swift/XRT/UVOT, ASAS-SN, and Fermi-LAT. We investigate the source's temporal behavior, identifying variability patterns and spectral changes in $\gamma$-ray emission that provide insights into the underlying physical processes. Additionally, we also search for a transient quasi-periodic oscillation in a 10-day binned $\gamma$-ray light curve. By employing the one-zone leptonic model, we fitted the broadband spectrum in different flux states and constrained the underlying physical parameters in different flux states. 
This work is structured in the following ways: \S\ref{introduction} introduction of the Source. Section \S\ref{sec:1} details observations and data processing techniques. In section \S\ref{temp:2}, we report the multi-wavelength temporal and spectral results of S5\,1803+784 using Swift/UVOT/XRT, ASAS-SN, and Fermi-Lat data. Finally, the paper concludes with a summary and discussion in Section \S\ref{sum}.

\section{Observations and Data reduction}
\label{sec:1}

\subsection{\textbf{Swift-XRT}}
Swift is a multi-wavelength space-based observatory equipped with  Burst Alert Telescope (BAT), X-ray Telescope (XRT), and Ultraviolet/Optical Telescope (UVOT) \citep{2005SSRv..120..165B}. Covering a broad spectrum of electromagnetic radiation, the Swift observes the sky across multiple wavebands, including Optical, Ultraviolet, soft, and hard X-rays. The Swift XRT is sensitive to soft X-rays in the energy range of 0.3 -- 10 keV, it helps in measuring the fluxes, light curves, and spectra of the source. Swift conducted a total of 43 observations of the source S5\,1803+784, these observations include both the quiescent and flaring states. Using \emph{“xrtpipeline\footnote{\url{https://www.swift.ac.uk/analysis/xrt/xrtpipeline.php}}”} Version: 0.13.5, and the calibration file (CALDB, version: 20190910), we created the cleaned event files for the photon counting (PC) mode of the data. The “xselect” (v2.5b) tool \footnote{\url{https://heasarc.gsfc.nasa.gov/docs/software/lheasoft/ftools/xselect/index.html}} was used to extract the source and background spectrum files. 
A circular region with a radius of 25 pixels was centered on the source location to define the source region, while a background region was selected as a circular area offset by 50 pixels from the source.
To create the required auxiliary response file (ARF),  “xrtmkarf\footnote{\url{https://www.swift.ac.uk/analysis/xrt/arfs.php}}” tool was used. The “grppha\footnote{\url{https://heasarc.gsfc.nasa.gov/ftools/caldb/help/grppha.txt}}” task was applied to ensure the validity of the spectrum for $\chi^2$ statistics. This involved binning the spectrum to ensure a minimum of 20 counts per bin. The HEASOFT package "XSPEC\footnote{\url{https://heasarc.gsfc.nasa.gov/xanadu/xspec/}}" (v12.13.0c)  \citep{1996ASPC..101...17A} was used to perform X-ray spectra analysis in the energy range 0.3 -- 10.0\, keV. The X-ray spectra were fitted with a Tbabs $\times$ power law (PL)/broken power law (BPL)/log parabola model. However, power law and log parabola provide a better fit during the X-ray spectral fit.
During the spectral fitting, the neutral hydrogen column density for this source was fixed at $N_H = 3.64 \times 10^{20}, \text{cm}^{-2}$ \citep{1996ApJS..105..369M}, while other parameters of the model were allowed to vary freely.\\

\subsection{\textbf{Swift-UVOT}}
%\subsection{\emph{Swift}-UVOT}
Swift-UVOT offers Optical/UV data through six filters U (3465 \AA), V (5468 \AA), B (4392 \AA), UVW1 (2600 \AA), UVM2 (2246 \AA), and UVW2 (1928 \AA) \citep{roming2005}. Swift-UVOT telescope \citep{2005SSRv..120...95R} has also observed the BL\,Lac source S5\,1803+784 simultaneously with the \emph{Swift}-XRT. The Swift-UVOT data for the source covers all six filters during epoch S4 but exhibits limited filter availability during epochs S1 (U, UVW1, UVW2), S2 (UVW1, UVM2), and S3 (U, UVW1, UVW2, UVM2).  We obtained data of S5 1803+784 from the HEASARC Archive and utilized HEASoft\footnote{\url{https://heasarc.gsfc.nasa.gov/docs/software/heasoft/}} (v6.26.1) to process it into a usable scientific format. The \textit{UVOTSOURCE}\footnote{\url{https://heasarc.gsfc.nasa.gov/lheasoft/help/uvotsource.html}} package included in HEASoft (v6.26.1)  was used for image processing, while the \textit{UVOTIMSUM}\footnote{\url{https://www.swift.ac.uk/analysis/uvot/image.php}} package combined multiple images across filters. Photometry adhered to the procedures outlined by \citet{poole2008}, with a circular source region of 5 arcsecs centered at the source location and a background region two times larger than the source region. The \texttt{uvotsource} tool was utilized to obtain the magnitude, which was then adjusted for reddening and galactic extinction using a $E(B-V)$ value of 0.0448 \cite{2011AAS...21831803S}. Subsequently, these corrected magnitudes
were converted into flux values using the zero point magnitudes
specific to Swift-UVOT, as described by \citet{giommi2006swift}.

\subsection{\textbf{Fermi-LAT}}
\emph{Fermi}-LAT serves as one of the two instruments onboard Fermi $\gamma$-ray space telescope, featuring a wide field of view of approximately 2.4 Sr \citep{2009ApJ...697.1071A}. It has functioned as a pair conversion telescope in near-Earth orbit since June 2008, capable of detecting high-energy $\gamma$-rays within the energy range of 20 MeV to 1 TeV \cite{2009ApJ...697.1071A}. Functioning in all-sky scanning mode, it systematically scans the entire sky every three hours, enabling the identification of numerous $\gamma$-ray sources. One of the sources that \emph{Fermi}-LAT frequently monitored is S5\,1803+784 and has been continuously observed since August 4, 2008, at 15:43:36 UTC. Using Fermitools version 2.2.0, the data was processed following the standard analysis specified in the \emph{Fermi}-LAT documentation\footnote{\url{https://fermi.gsfc.nasa.gov/ssc/data/analysis/documentation/}} \cite{2017ICRC...35..824W}.  A circular region of radius $15 ^{\circ}$  centered on the source location (RA: 270.173, Dec: 78.467) was used to extract events. Following the recommendations in the Fermitools document, the data were filtered using “evclass = 128” and “evtype = 3”. In the analysis, we have used the most recent instrument response function (IRF), \emph{“P8R3 SOURCE V3”}. In order to reduce contamination from $\gamma$-rays originating from the Earth limb, a zenith angle cut greater than $90^{\circ}$  is applied. Two models were used to produce XML files: the isotropic background model “$\rm iso_-P8R3_-SOURCE_-V3_-v1.txt$\footnote{\label{fermi} \url{https://fermi.gsfc.nasa.gov/ssc/data/access/lat/BackgroundModels.html}}” and the galactic diffusion model “$\rm gll_-iem_- v07.fits$\footref{fermi}”.\par
We adopted criteria to filter out the sources with low Test Statistics (TS) i.e. below TS = 9 and generated a 10-day binned $\gamma$-ray light curve of the source of interest with TS$\> (\ge 9)$. In this process, parameters for sources positioned beyond 15$^\circ$ from the center of ROI were fixed, while for $\le 15^\circ$ were left free and were allowed to vary freely. In this project, We used the \texttt{FERMIPY}\footnote{\url{https://fermipy.readthedocs.io/en/latest/}} package to generate the 10-day binned $\gamma$-ray light curve.
 
\subsection{\textbf{ASAS-SN}} 
All-Sky Automated Survey for Supernovae (ASAS-SN; \cite{shappee2014man, kochanek2017all} is a global network of 24 telescopes that has been monitoring the extragalactic sky since 2012. With a limiting magnitude of $\sim$17.5-18.5 in the g-band and $\sim$16.5-17.5 in the V-band depending on lunation, ASAS-SN provides valuable observations across the entire sky. For this study, we utilized g-band data, Figure \ref{Fig-MWLight curve} collected through the ASAS-SN Sky Patrol (V2.0\footnote{\url{http://asas-sn.ifa.hawaii.edu/skypatrol/}}; \cite{shappee2014man, hart2023asas}.

\section{Multi-wavelength Temporal  and  Spectral results }
\label{temp:2}
\begin{figure*}
    \centering
    \includegraphics[width=0.9\textwidth]{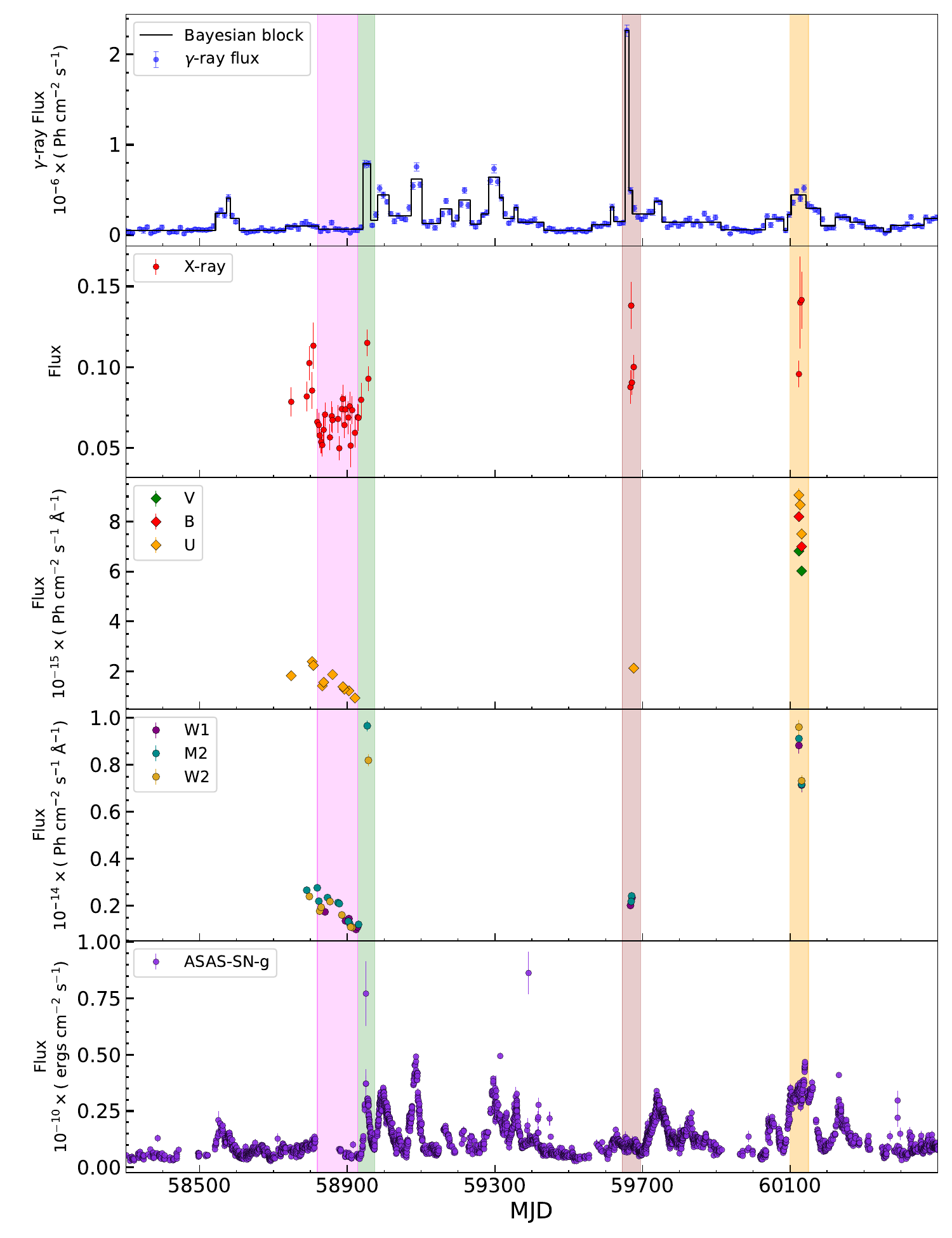}
    \caption{Multi-wavelength light curve (MWLC) obtained from ASAS-SN, Swift-XRT/UVOT, and Fermi-LAT. The top panel displays a 10-day binned $\gamma$-ray light curve (blue) obtained in the energy range of 100 MeV to 300 GeV along with the Bayesian Block (black). The top second is the light curve of X-ray data using Swift/XRT. The third and fourth panels depict Optical/UV light curves, where each point corresponds to a single observation ID. The last panel represents the ASAS-SN g-band light curve. The shaded regions in pink, green, brown, and orange throughout all five panels represent the S1, S2, S3, and S4 selected observation epochs, respectively. }
    \label{Fig-MWLight curve}    
\end{figure*}

To investigate the temporal behavior of LSP blazar S5 1803+784, Swift-XRT/UVOT observed 43 observations between MJD 58748 -- 60152 along with Fermi-LAT
. The X-ray and optical/UV data were analyzed using the methods outlined in \S\ref{sec:1}. Figure \ref{Fig-MWLight curve} displays the resulting multi-wavelength light curves, including 10-day binned $\gamma$-ray observations in the top panel of Figure \ref{Fig-MWLight curve} along with the Bayesian Blocks. The X-ray, optical, and UV data are given in the second, third, and fourth panels from the top, respectively. The bottom panel of Figure \ref{Fig-MWLight curve} presents the ASAS-SN simultaneous observations. \par
%\st{Figure 1 presents the resulting light curves, with the X-ray data in the second panel and the optical/UV data in the bottom panel. Each point on the light curves represents a single observation. In the $\gamma$-ray light curve, we utilized the 3-day binned light curve from the Fermi-LAT Light Curve Repository, applying a minimum detection significance threshold (TS = 9). The flux values were integrated over the energy range of 0.1 - 100 GeV by} \cite{abdollahi2023fermi}.\par

The identification of flare events in the $\gamma$-ray light curve lacks a universally accepted method. However, the Bayesian Blocks (BB) algorithm has gained some popularity for detecting flares in AGN light curves. In this study, we employed the Bayesian Blocks (BB) algorithm\footnote{\url{https://docs.astropy.org/en/stable/api/astropy.stats.bayesian_blocks.html}} of the Astropy package, setting the false-alarm probability to $p_0 = 0.05$, Figure \ref{Fig-MWLight curve}. While applying the BB algorithm, we excluded upper limits from the $\gamma$-ray light curve analysis, as the algorithm is well-suited for handling non-uniform light curves. To estimate the flaring temporal regions, we utilized the HOP algorithm, defining a flare as any event satisfying the condition $F_{\text{BB}} > 3 \times \overline{F}$, where $\overline{F} = 1.152\times 10^{-7} ph \ cm^{-2} \ s^{-1}$. Accordingly, we identified four epochs having simultaneous multi-wavelength observations, in which epoch S1 represents a low flux state and epochs S2, S3, and S4 represent three flaring states, see the Figure \ref{Fig-MWLight curve}.\par    

The average $\gamma$-ray flux was %$1.95\times10^{-07} ph \ cm^{-2} \ s^{-1}$ 
$6.05\times10^{-08} ph \ cm^{-2} \ s^{-1}$ during 58819 – 58928 (S1). A  high-flux state (S2) occurred during MJD %58959 -- 58974
58928 -- 58974, characterized by an increased average $\gamma$-ray flux of %$2.17\times10^{-07} ph \ cm^{-2} \ s^{-1}$
$4.40\times10^{-07} ph \ cm^{-2} \ s^{-1}$
. Similarly, a flux state (S3) occurred during MJD 59645 - 59695, with  average
$\gamma$-ray flux of %$1.19\times10^{-06} ph \ cm^{-2} \ s^{-1}$
$6.78\times10^{-07} ph \ cm^{-2} \ s^{-1}$. This was followed by another state (S4), identified during MJD 60100 -- 60150, with an average $\gamma$-ray flux of %$3.83\times10^{-07} ph \ cm^{-2} \ s^{-1}$
$4.19\times10^{-07} ph \ cm^{-2} \ s^{-1}$. A comprehensive multi-wavelength analysis of an LSP blazar was conducted, covering four epochs (S1, S2, S3, S4) and multiple energy bands. The study revealed complex variability patterns across the four epochs, with each epoch displaying distinct behaviors in three energy bands. Specifically, epoch S1 showed consistently low flux levels, while epoch S2 displayed high flux levels across all bands. In contrast, epoch S3 exhibited high $\gamma$-ray and X-ray flux, but low UVOT flux, and epoch S4 was characterized by the elevated flux in UVOT, X-ray, and $\gamma$-ray bands, offers clues to understanding the elusive nature of orphan blazars.

%\begin{figure*}
%\centering

%\includegraphics[scale=0.5]{draft_lc.eps}\\

%\caption{Multi-wavelength light curve (MWLC) obtained through observations from  Swift-XRT/UVOT, and Fermi-LAT.The top panel displays a 3-day binned $\gamma$-ray light curve obtained in the energy range of 0.3 to 300 GeV. The top second is the light curve of X-ray data using Swift/XRT. The last two panels depict Optical/UV light curves, where each point corresponds to a single observation ID.}
%\label{fig:mwl}
%\end{figure*}

\subsection{\textbf{Fractional Variability}}
To comprehensively study the statistical variability characteristic of the blazar S5 1803+784, we utilized the simultaneous multi-wavelength observations from Fermi-LAT, Swift-XRT/UVOT, and ASAS-SN, following the approach described in \cite{Edelson_2002, 2003MNRAS.345.1271V}. \par

To measure the variability in a given sample of $\mathcal{N}$ data points $x_{i}$, the variance $\mathcal{S}^2$ is used, defined as:

\begin{equation}
\mathcal{S}^2 = \frac{1}{\mathcal{N}-1}\sum_{i=1}^{\mathcal{N}}(\langle x\rangle-x_{i})^{2}
\end{equation}

In the case of blazars, the light curve is the sample, which consists of measurement uncertainties $\sigma_{\text{err},i}$. The fractional variability tool is considered to account for these uncertainties. It incorporates the variance of the data points and the cumulative variance that is created from these measurement uncertainties. This is commonly called fractional variance excess or normalized excess variance, given by \cite{Edelson_2002}.
To measure the variability of a source in different energy bands, the fractional variability amplitude expression has been used and is given by \cite{2003MNRAS.345.1271V}:

\begin{equation}
    \label{eq3}
    F_{\text{var}} = \sqrt{\frac{S^2 - \overline{\sigma_{\text{err}}^2}}{\overline{F}^2}}
\end{equation}

where $S^2$ is the variance, $\overline{F}$ is the mean, and $\overline{\sigma_{\text{err}}^2}$ is the mean square of the measurement error on the flux points. The uncertainty on $F_{\text{var}}$ is given by \cite{2003MNRAS.345.1271V}:

\begin{equation}
    \label{eq4}
    F_{\text{var,err}} = \sqrt{\frac{1}{2N}\left(\frac{\overline{\sigma_{\text{err}}^2}}{F_{\text{var}}\overline{F}^2}\right)^2 + \frac{1}{N}\frac{\overline{\sigma_{\text{err}}^2}}{\overline{F}^2}}
\end{equation}

Here, N represents the number of flux points in the light curve. \\

\begin{figure}
\centering
\includegraphics[width=0.47\textwidth]{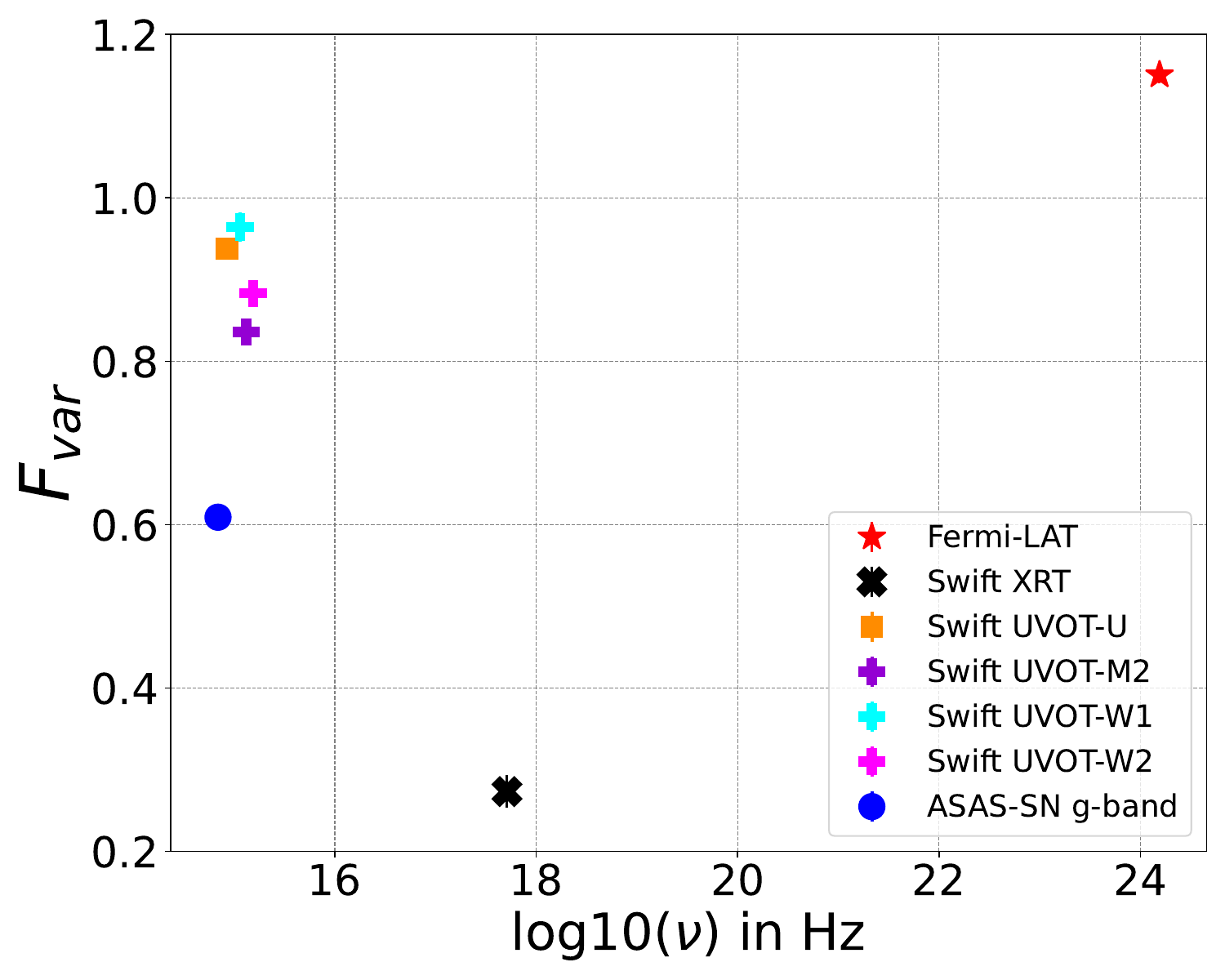}
\caption{Fractional variability amplitude obtained in different energy bands is plotted against the frequency }

\label{fig:Fvar_MWL}
    
\end{figure}

For the $\gamma$-ray, swift X-ray, swift-Optical/UV, and ASAS-SN light curves, the values of $F_{\text{var}}$ and their uncertainties are shown in Table \ref{tab:4}.  Figure~\ref{fig:Fvar_MWL} shows the relation between $F_{\text{var}}$ and frequency. The X-ray band has a lower variability amplitude value when compared to the values seen in the $\gamma$-ray, optical, and UV bands. Remarkably, it represents a distinct pattern to what we have observed in the BL\,Lac class of earlier sources such as Mrk\,501 and Mrk\,421 i.e., higher in X-rays compared to $\gamma$-rays \cite{2019Galax...7...62S,abe2023multi, Abe:2024eba, Tantry:2024qvs}. The $F_{var}$ displays a double-peak structure with higher variability at UV/Optical and $\gamma$-ray from the 4-year data set. Previous studies by \cite{2020A&A...634A..80R} utilized  $\gamma$-ray Space Telescope data to analyze monthly $\gamma$-ray flux variability in blazars; the dataset comprises 1120 blazars, including 481 FSRQs and 639 BL\,Lacs. Mean  $F_{var}$ of 0.55 ± 0.33 for FSRQs, while for LSP blazars  mean $F_{var}$ value of 0.54 ± 0.32. Furthermore, we quantify the variability of the $\gamma$-ray light curve by comparing the fractional amplitude values between low and high flux states, using $\sim$15 yr long lightcurve with the binning of 3-day and 7-day, as illustrated in Figure\ref{fig:fv_plot}.

%%\begin{table}
%%\centering
%%\caption{Fractional variability amplitude $F_{var}$ obtained in different energy bands. Columns 1: energy band 2: fractional variability amplitude.}
%%\begin{tabular}{l r}

%%\multicolumn{2}{c}{\textbf{Data set (2019--2023)}}\\
%Data set(2019-2023)&\\
%%\hline
%%Energy band  & $\rm F_{var}$\\
%%\hline
%%$\gamma$-ray (\mbox{0.1 -- 300\,GeV}) & 0.92$\pm$0.04\\  
%%X-ray (0.3--10 keV) & 0.29$\pm$0.02  \\
%%UVW2 & 0.59$\pm$0.02 \\
%%UVM2 & 0.82$\pm$0.03\\
%%UVW1 & 0.94$\pm$0.01\\
%%U & 0.94 $\pm$ 0.02\\
%%\hline
%%\end{tabular}
%%\label{tab:1}
%%\end{table}

\begin{table}
\setlength{\extrarowheight}{4pt}
\setlength{\tabcolsep}{10pt}
\centering
\caption{Fractional variability amplitude $F_{var}$ obtained in different energy bands. Columns 1: energy band 2: fractional variability amplitude with uncertainty.}

\begin{tabular}{l c}

\multicolumn{2}{c}{\textbf{Data set (2019--2023)}}\\
\hline
Energy band  & $\rm F_{var}$\\
[+2pt]
\hline
Fermi $\gamma$-ray (\mbox{0.1 -- 300\,GeV}) & 1.15$\pm$0.01\\  
Swift X-ray (0.3--10 keV) & 0.27$\pm$0.02  \\
Swift UVOT-W2 & 0.88$\pm$0.01 \\
Swift UVOT-M2 & 0.83$\pm$0.009\\
Swift UVOT-W1 & 0.96$\pm$0.01\\
Swift UVOT-U & 0.93 $\pm$ 0.01\\
ASAS-SN g-band & 0.60$\pm$0.001 \\
[+4pt]
\hline
\end{tabular}
\label{tab:Fvar}
\end{table}

%\begin{figure}
%\centering
%\includegraphics[width=0.48\textwidth]{frac_var.eps}
%\caption{Fractional variability amplitude obtained in different energy bands is plotted against the energy.}

%\label{fig:gammavar}
    
%\end{figure}

\subsection{\textbf{Transient quasi-periodic oscillation}}
To investigate the presence of a transient Quasi-Periodic Oscillation (QPO) in the gamma-ray emission of the blazar S5\,1803+784, we employed several analytical approaches, including the Lomb-Scargle periodogram (LSP), the Weighted Wavelet Z-transform (WWZ), a first-order autoregressive AR(1) process, the Date-Compensated Discrete Fourier Transform (DCDFT), and Damped Random Walk (DRW). Below, we provide a detailed description of these methods and the results obtained from the analysis.\par

\subsubsection{\textbf{Lomb-Scargle Periodogram}}\label{sec:lsp}
The Lomb-Scargle periodogram is a widely used and popular technique for detecting periodic signals in time series data with unevenly spaced observations \cite{lomb1976least, scargle1982studies}. It is particularly well-suited for analyzing non-uniformly sampled light curves to identify potential periodicities. In our study, we utilized the \textsc{LOMB-SCARGLE}\footnote{\url{https://docs.astropy.org/en/stable/timeseries/lombscargle.html}} class from the \textsc{ASTROPY} library to perform the analysis. Additionally, the uncertainties associated with the flux measurements were incorporated into the analysis. The periodogram function of the Lomb-Scargle method is defined as \cite{vanderplas2018understanding}\par

\begin{equation}
\begin{split}
P_{LS}(f) = \frac{1}{2} \bigg[ &\frac{\left(\sum_{i=1}^{N} x_i \cos(2\pi f (t_i - \tau))\right)^2}{\sum_{i=1}^{N} \cos^2(2\pi f (t_i - \tau))} \\
&+ \frac{\left(\sum_{i=1}^N x_i \sin(2\pi f (t_i - \tau))\right)^2}{\sum_{i=1}^N \sin^2(2\pi f (t_i - \tau))} \bigg]
\end{split}
\end{equation}

where, $\tau$ is 
\begin{equation}
    \tau = \rm{tan^{-1}} \left(\frac{\sum_{i=1}^N sin\left(2\pi f (t_i - \tau) \right)}{2 (2\pi f) \sum_{i=1}^N cos\left(2\pi f (t_i - \tau) \right)} \right)
\end{equation}
In our analysis, we selected the minimum ($f_{min}$ and maximum ($f_{max}$) temporal frequencies as 1/T and 1/2$\Delta T$, respectively. Here, T represents the total observation duration, and $\Delta T$ denotes the binning interval of the light curve or the time difference between consecutive data points. These parameters are critical for the Lomb-Scargle periodogram (LSP) analysis. The analysis identified a potential peak corresponding to a frequency of $ 0.00243 \pm 0.00045 \ day^{-1} $ ($\sim$411 days). The uncertainty in the detected period was estimated by fitting a Gaussian function to the periodogram peak, with the half-width at half-maximum (HWHM) of the Gaussian taken as the error on the observed period \cite{vanderplas2018understanding, sharma2024detection}. \par

\subsubsection{\textbf{Weighted Wavelet Z-Transform}}\label{sec:wwz}
In addition, Wavelet analysis is a robust technique for detecting periodic signals in time series by simultaneously decomposing them into frequency and time domains. This method is particularly effective for studying the evolution of QPO features, as it provides detailed insights into how potential periodic signals emerge, evolve, and fade over time \cite{foster1996wavelets}.\par
In wavelet analysis, we used the abbreviated Morlet kernel, which has the following functional form:

\begin{equation}
    f[\omega (t - \tau)] = \exp[i \omega (t - \tau) - c \omega^2 (t - \tau)^2]
\end{equation}
and the corresponding WWZ map is given by,
\begin{equation}
    W[\omega, \tau: x(t)] = \omega^{1/2} \int x(t)f^* [\omega(t - \tau)] dt
\end{equation}
Here, $f^*$ is the complex conjugate of the wavelet kernel f, $\omega$ is the frequency, and $\tau$ is the time-shift. For more details, see \cite{sharma2024detection}, and references therein. In this study, we employed the publicly available Python WWZ\footnote{\url{https://github.com/eaydin/WWZ}} code to search for periodicities. The resulting WWZ map revealed a distinct concentration of power around a frequency $0.00251\pm 0.0002 \ day^{-1}$ ($\sim$ 399 days). The uncertainty on the observed period was estimated by fitting the Gaussian function to the observed prominent peak in the average WWZ. \par

\subsubsection{\textbf{First order Autoregressive process (REDFIT)}}\label{sec:redfit}
AGN light curves are generally dominated by red noise, which arises from stochastic processes in the accretion disc or jet. The emissions from blazars are well modeled using the simplest autoregressive (AR) processes, known as a first-order autoregressive (AR(I)) process \cite{schulz2002redfit}, where the emission at an instance is related to past activities, defined as, $r(t_i)=A_i r(t_{i-1})+\epsilon(t_i)$, where the current emission ($r_t$) depends linearly on the previous emission ($r_{t-1}$), $A_i = exp\left( -(t_i - t_{i-1})/\tau \right) \in [0,1]$, A is the average autoregressive coefficient calculated from the mean of the sampling intervals, $\tau$ is the timescale of autoregressive process, and a random error ($\epsilon_t$). The power spectrum of an AR(I) process is defined as:

\begin{equation}
    G_{rr}(f_i) = G_0 \frac{1 - A^2}{1 - 2 A cos\left( \frac{\pi f_i}{f_{Nyq}} \right) + A^2}
\end{equation}

where $G_0$ is the average spectral amplitude, $f_i$ are the frequencies, and $f_{Nqy}$ is the Nyquist frequency. In this study, we used the publicly available r programming code REDFIT\footnote{\url{https://rdrr.io/cran/dplR/man/redfit.html}} to estimate the red noise-corrected power spectrum. We calculated the significance of the powers in the power spectrum using equation (6). The obtained power spectrum clearly shows a significant peak at frequency $0.00237\pm 0.00049 \  day^{-1}$ ($\sim421 \ days$), exceeding 99$\%$ significance level. The uncertainty value in the dominant peak frequency is HWHM from the fitting of Gaussian to the peak profile in the REDFIT power spectrum. \par

\subsubsection{\textbf{Date-compensated Discrete Fourier Transform}}\label{sec:dcdft}
In this study, we also utilized the date-compensated discrete Fourier transform (DCDFT) \citep{ferraz1981estimation, foster1995cleanest, fan2007radio}, a Fourier-based technique. To identify periodic signatures in an unevenly sampled time series, we applied a least-squares regression using three trial functions: $\phi_1$(t)=1, $\phi_2$(t)=cos($\omega$t), and $\phi_3$(t)=sin($\omega$t).\par
The observations x(t$_i$) are defined as data vector

\begin{equation}
    \ket{x}=\left[ x(t_1), x(t_2), ..., x(t_N)  \right]
\end{equation}

To project the data onto the subspace spanned by the three trial functions defined above, we construct a set of trial vectors based on these functions, which is given as 

\begin{equation}
    \ket{\phi_{\alpha}}=\left[ \phi_{\alpha}(t_1), \phi_{\alpha}(t_2), ..., \phi_{\alpha}(t_N)  \right]
\end{equation}

where, $\alpha$=1,2,3.
The projection of $\ket{x}$ onto the subspace spanned by $\ket{\phi_{\alpha}}$ yields a model vector $\ket{y}$ and a residual vector $\ket{\Theta}$,

\begin{equation}
    \ket{x}=\ket{y}+\ket{\Theta}
\end{equation}

here, the model vector is defined as

\begin{equation}
    \ket{y}=\sum_{\alpha} c_{\alpha}\ket{\phi_{\alpha}},
\end{equation}

while the residual vector is orthogonal to the subspace, 

\begin{equation}
    \langle \phi_{\alpha} | \Theta \rangle=0, \ \ \rm{for} \ \  \rm{all} \ \ \alpha
\end{equation}

now, taking the inner product of the trial vector with data vector, we can obtain coefficient $c_{\alpha}$,

\begin{equation}
    \langle \phi_{\alpha} | x \rangle=\sum_{\beta} c_{\beta} \langle \phi_{\alpha} |  \phi_{\beta} \rangle = \sum_{\beta} S_{\alpha \beta} c_{\beta}
\end{equation}

where, $S_{\alpha \beta}$ defines the metric tensor for the trial subspace. The coefficient $c_{\alpha}$ can be obtained by multiplying $S_{\alpha \beta}^{-1}$ in both sides in equation 14 

\begin{equation}
    c_{\alpha} = \sum_{\beta} S_{\alpha \beta}^{-1}\langle \phi_{\beta} | x \rangle
\end{equation}

\begin{figure}
    \centering
    \includegraphics[width=0.47\textwidth]{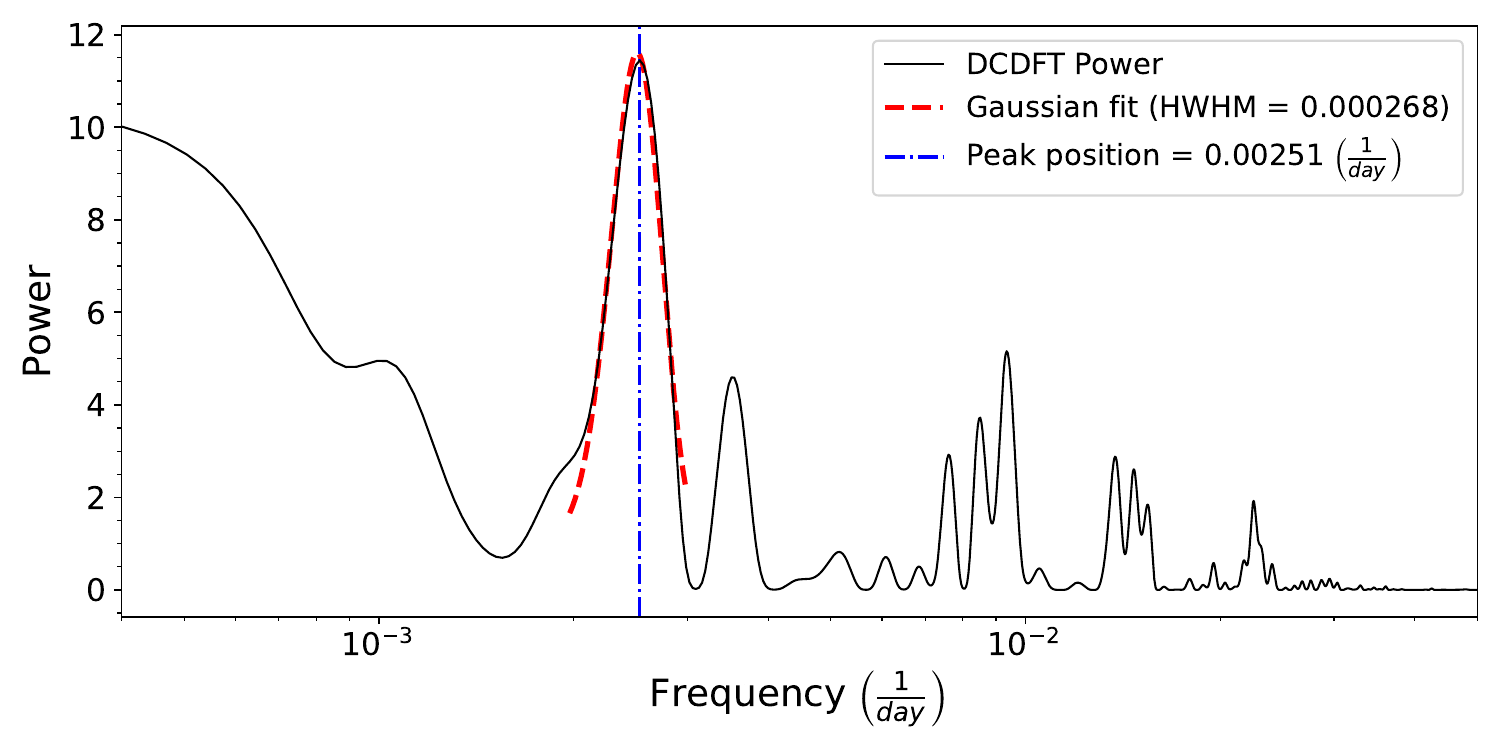}
    \caption{The date-compensated discrete Fourier transform (DCDFT) power level of $\gamma$-ray light curve of S5 1803+784 is presented here. This figure highlights the dominant peak (black) at 0.00251$\pm$0.000268 $\rm{day^{-1}}$ (blue vertical line). To estimate the uncertainty on the period, the peak is fitted with a Gaussian function (dashed red curve), and the half-width at half maximum is used as the uncertainty measure.}
    \label{Fig-DCDFT}    
\end{figure}

Finally, the power level of DCDFT is defined as

\begin{equation}
    P_X(\omega)=\frac{1}{2} N\left[ \langle y | y \rangle - \langle 1 | y \rangle^2  \right]/s^2
\end{equation}

where $s^2$ is the flux variance. We applied this procedure to estimate the power level of DCDFT, see Figure \ref{Fig-DCDFT}. The dominant peak in DCDFT power spectrum is detected at 0.00251$\pm$0.000268 $\rm{day^{-1}}$ ($\sim$398 days). The uncertainty on this peak is determined using a method similar to that employed in LSP, WWZ, and REDFIT. The results, presented in Table \ref{tab:QPO_all}, are consistent with each other, including their respective uncertainties.

\subsubsection{\textbf{Damped Random Walk Model}}\label{sec:drw}
The observed variability in AGNs emission is stochastic in nature and can be well described by the simplest model of the Continuous Autoregressive Moving Average [CARMA(p,q)] \citep{kelly2009variations}, known as Damped Random Walk (DRW) Model. This is a widely accepted model to characterize the red noise behavior in emissions of AGNs, which is defined as the solutions to the following differential equation \citep{kozlowski2009quantifying, macleod2012description, ruan2012characterizing, zu2013quasar, moreno2019stochastic, burke2021characteristic, zhang2022characterizing, zhang2023gaussian, sharma2024probing, zhang2024discovering, sharma2024microquasars}:

\begin{equation}
    \left[ \frac{d}{dt} + \frac{1}{\tau_{DRW}} \right] y(t) = \sigma_{DRW} \epsilon(t)
\end{equation}

where $\tau_{DRW}$  and $\sigma_{DRW}$ are the characteristic damping time-scale and amplitude of the DRW process, respectively. The covariance function of the DRW model is defined as
%\begin{equation}
%    k(t_{nm}) = a.exp(-t_{nm}c),
%\end{equation}

\begin{equation}
\centering
    k(t_{nm}) = a \cdot \exp(-t_{nm} \, c),
\end{equation}

where $t_{nm} = | t_n -t_m|$ denotes the time lag between measurements m and n, with $a = 2 \sigma_{DRW}^2$ and $c = \frac{1}{\tau_{DRW}}$. The power spectral density (PSD) of DRW model is defined as:
\begin{equation}
    S(\omega) = \sqrt{\frac{2}{\pi}} \frac{a}{c} \frac{1}{1 + (\frac{\omega}{c})^2}
\end{equation}

The DRW PSD  has a form of Broken Power Law (BPL), where the broken frequency $f_b$ corresponds to the characteristic damping timescale $\tau_{DRW} = \frac{1}{2\pi f_b}$.
In the modeling process, we utilized a publicly available Python package \textsc{EzTao}\footnote{\url{https://eztao.readthedocs.io/en/latest/}}, which is built on top of the \textsc{celerite}\footnote{\url{https://celerite.readthedocs.io/en/stable/}} package. In the parameters estimation, we employed the Markov Chain Monte Carlo (MCMC) algorithm, which is implemented in the \textbf{\texttt{emcee}}\footnote{\url{https://github.com/dfm/emcee}} package. In MCMC modeling, a total of 25,000 samples are generated, and the first 10,000 samples are considered burn-in. From the remaining MCMC samples, we calculated the values of model parameters and their uncertainties.

\begin{figure*}
    \centering
    \includegraphics[width=0.85\textwidth]{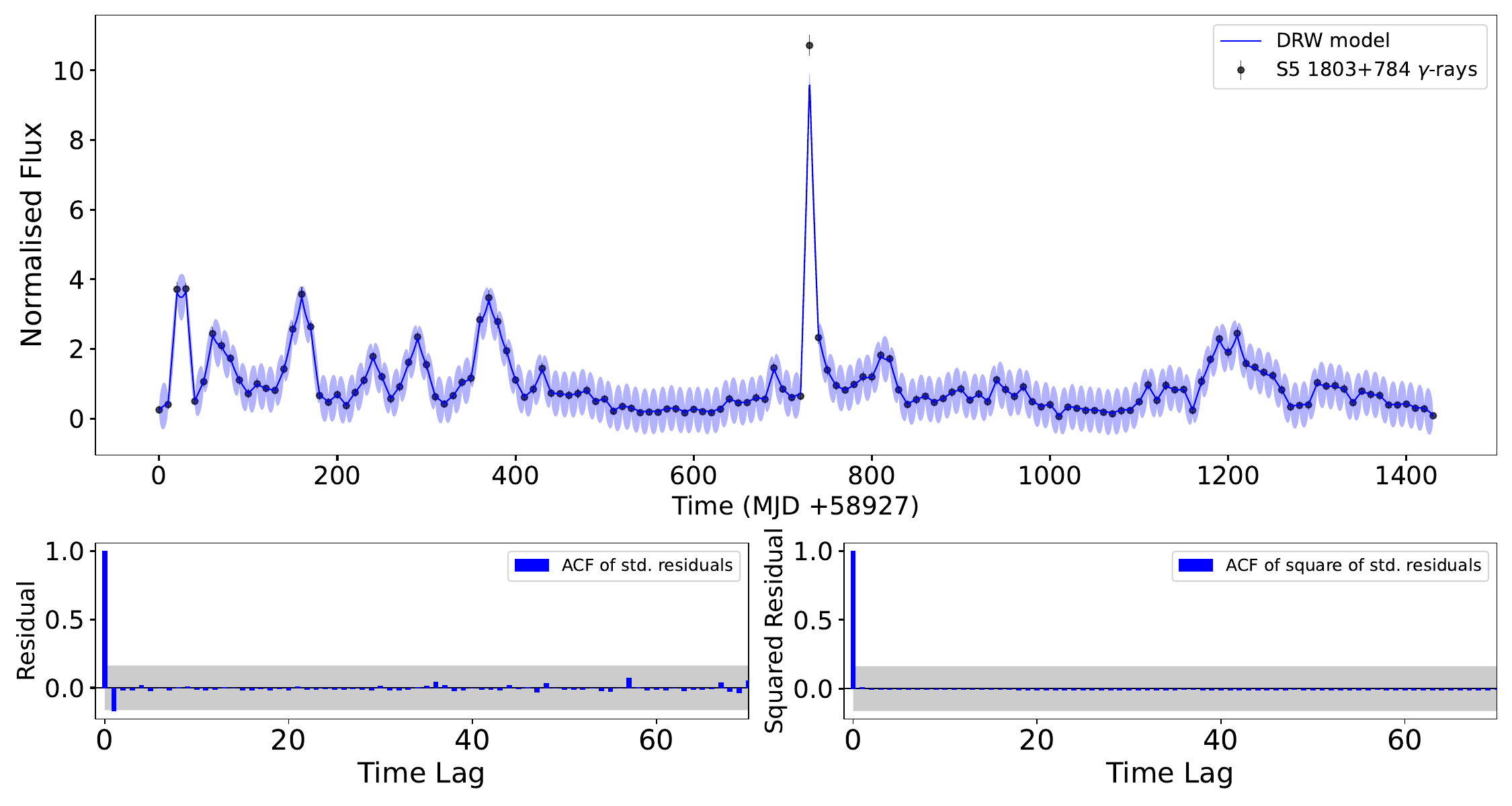}
    \caption{The DRW modeling of a 10-day binned $\gamma$-ray light curve of blazar S5 1803+784 over 1400 days from the time stamp MJD 58927. This figure shows the output of modeling, the top panel represents the $\gamma$-ray flux points with their uncertainties, with the best-fit DRW model profile in blue, along with the 1$\sigma$ confidence interval. The bottom panels represent the autocorrelation functions (ACFs) of the standardized residuals (bottom left) and the squared of standardized residuals (bottom right), respectively, along with 95$\%$ confidence intervals of the white noise.}
    \label{Fig-DRW_LC}    
\end{figure*}

\begin{figure}
    \centering
    \includegraphics[width=0.49\textwidth]{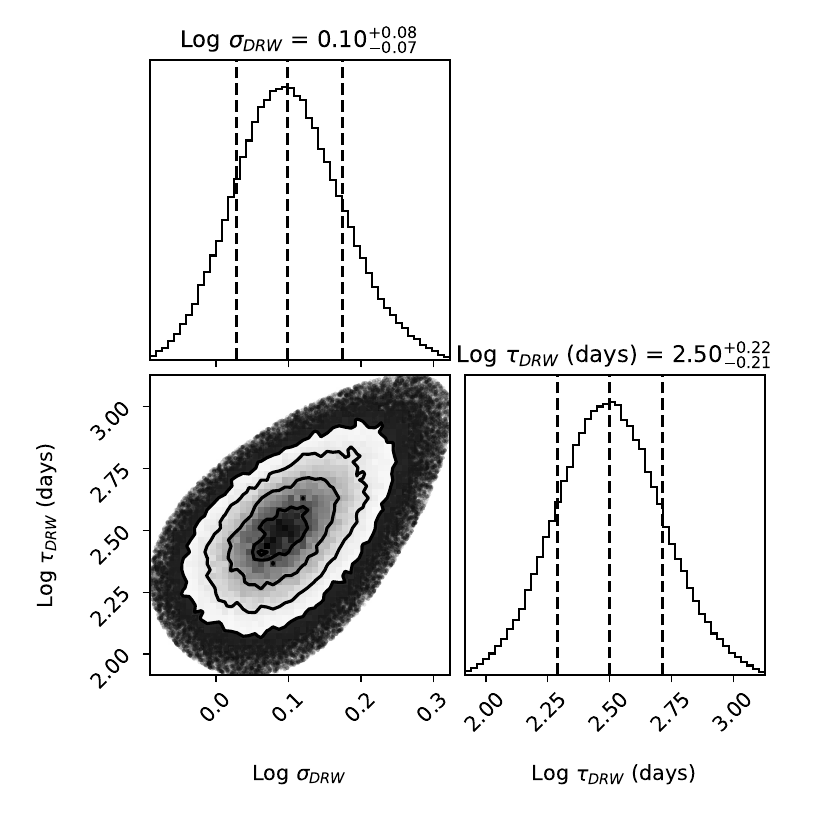}
    \caption{The posterior probability distributions of the DRW model parameters obtained from the modeling of $\gamma$-ray light curve of S5 1803+784 are shown here. }
    \label{Fig-Corner_plot}    
\end{figure}

The findings of the modeling of the $\gamma$-ray light curve with DRW model are tabulated in Table \ref{tab:DRW} and shown in Figures \ref{Fig-DRW_LC}, \ref{Fig-Corner_plot}. The fit quality is evaluated through the autocorrelation functions of standardized residuals and the square of standardized residuals, Figure \ref{Fig-DRW_LC}.\par 

The results from all three approaches are consistent with each other, considering their uncertainties, summarized in Table \ref{tab:QPO_all}.

\begin{table}
\setlength{\extrarowheight}{7pt}
\setlength{\tabcolsep}{8pt}
\centering
\caption{The observed DRW model parameters from the modeling of $\gamma$-ray light curve of blazar S5 1803+784 are tabulated here.}

\begin{tabular}{c c c }
\hline
\hline
Source  & $\rm{Log} \ \sigma_{DRW}$ & $\rm{Log} \ \tau_{DRW} \ (\rm{days})$ \\
(1) & (2) & (3) \\
[+2pt]
\hline
S5 1803+784 &  0.10$_{-0.07}^{+0.08}$ & 2.50$_{-0.21}^{+0.22}$ \\  
[+5pt]
\hline
\end{tabular}

\label{tab:DRW}
\end{table}

\subsection{\textbf{Significance estimation}}
\label{sec:Significance}
As we discussed, the red noise characteristic of AGN or Blazar's emissions is believed to originate from stochastic processes and is well represented by a power-law form defined as $P(\nu) \sim A \nu^{- \beta}$, where $\nu$ represents the temporal frequency and $\beta > 0$ represents the spectral slope. The best-fitting power-law model slope is $1.03\pm0.3$. It is crucial to estimate the significance of the dominant peaks obtained from the adopted techniques. In REDFIT analysis \cite{schulz2002redfit}, the significance was estimated from the $\chi^2$ distribution of the periodogram using the equation (8). \par
\begin{figure*}
    \centering
    \includegraphics[width=0.9\textwidth]{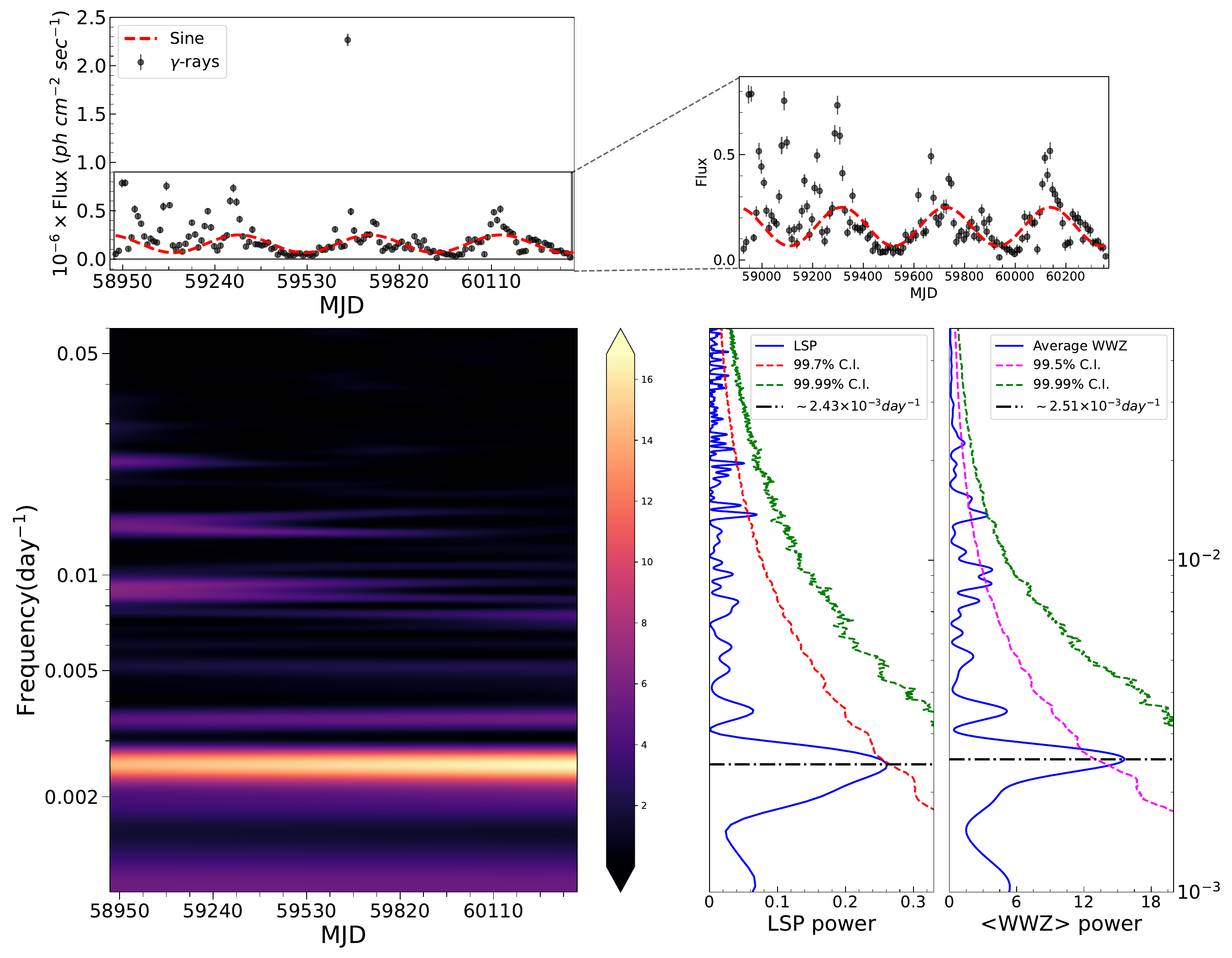}
    \caption{The $\gamma$-ray light curve is analyzed using the Lomb-Scargle Periodogram (LSP) and Weighted Wavelet Z-transform (WWZ) methods. The top-left panel displays the $\gamma$-ray light curve overlaid with a sinusoidal fit. A zoomed-in view of the light curve, excluding the highest flare at around $\sim$59675 MJD (only for clear visibility of the periodic signal), is shown in the top-right panel. The bottom panels present the WWZ map (left), LSP (middle), and average WWZ (right), respectively. The observed local significance of the peak of the timescale of $\sim$411 days from the LSP analysis is $99.7\%$ (red curve) and with a timescale of $\sim$399 days from WWZ analysis is exceeding $99.5\%$.}
    \label{Fig-LSP}    
\end{figure*}

Besides the above approach, we used a Monte Carlo simulation approach to estimate the significance of the LSP and WWZ dominant peaks. In this, we simulated the $1\times 10^5$ synthetic light curves, having the same PSD and PDF profiles as the original light curve, using the approach of \cite{emmanoulopoulos2013generating}. We estimated the local significance of observed peaks from the power spectrum distribution at each candidate frequency. In the LSP analysis, the observed peak at $\sim$0.00243 $day^{-1}$ (411 days) has a significance level of 99.7$\%$, while peak at $\sim$0.00251 $day^{-1}$ (399 days) exceeding 99.5$\%$ in WWZ analysis. These findings are consistent with REDFIT analysis and the observed dominant peak frequency is $\sim$0.00237 $day^{-1}$ (421 days with a significance level surpassing 99$\%$. The observed findings confirm the detection of a possible transient quasi-periodic signal in the $\gamma$-ray light curve (MJD 58900 -- 60350) with three complete cycles, see the top left panel of Figure \ref{Fig-LSP}.  \par

\begin{figure*}
    \centering
    \includegraphics[width=0.7\textwidth]{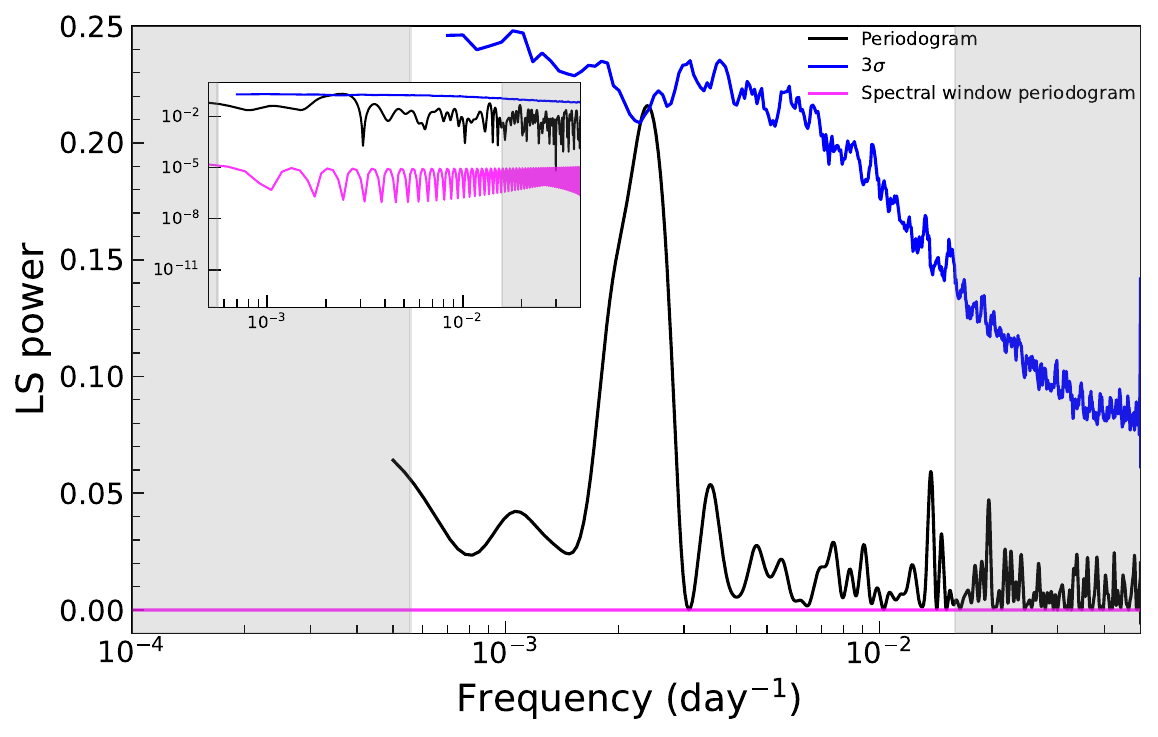}
    \caption{The LSP of the $\gamma$-ray light curve of the blazar S5 1803+784 is shown here. The figure includes the LS periodogram of the original light curves (black) and spectral window (pink). The significance level (blue) of the dominant observed peak in LSP is estimated using DRW mock light curves. The observed peak at 0.00243 $day^{-1}$ exceeds 3$\sigma$ significance level. A logarithmic version of the periodograms is provided in the sub-figure. }
    \label{Fig-LSP_DRW}    
\end{figure*}

In addition to the previously discussed methodologies for estimating the significance of the observed QPO signal in $\gamma$-ray emissions, we analyzed the $\gamma$-ray light curve using the DRW model. As mentioned earlier, the DRW model is a widely used red noise model capable of characterizing the variability in AGN emissions. Using the EzTao package, we simulated 15,000 mock light curves using the optimal amplitude and damping timescale values with a sampling rate consistent with the real observations. By utilizing the all mock light curves, we computed Lomb-Scargle periodograms for each. To estimate the significance level of the dominant peak in original $\gamma$-ray LSP, we calculated the 84th, 97.5th, 99.85th, and 99.995th percentiles of the 15000 mock LSPs for each candidate frequency value, which correspond to the 1$\sigma$, 2$\sigma$, 3$\sigma$, and 4$\sigma$ significance level. This analysis reveals that the peak at 0.00243 $day^{-1}$ exceeds the 3$\sigma$ significance level. This finding is consistent with the results from other methodologies. We also calculated the spectral window periodogram by constructing a light curve with a total number of time stamps ten times larger than the original one within the observed temporal frame. In this spectral window light curve, the time stamps matching the original observations are assigned a value of one, while all others are set to zero. Further, we applied the LSP method to generate the periodogram, as shown in pink in Figure \ref{Fig-LSP_DRW}. This analysis highlights that for unevenly sampled time series, the spectral window periodogram can exhibit spurious peaks that arise due to non-uniform sampling in the light curve, have power comparable to the original signal, aiding in the identification of false detections. In our case, the $\gamma$-ray light curve in the given time frame is well sampled and spectral window periodogram did not exhibit any false signals. However, in such studies the results can be influenced by the uneven sampling and finite length of the light curve. Previous studies \citep{kozlowski2017limitations, suberlak2021improving} have demonstrated potential biases in variability measurements due to these limitations. \citep{burke2021characteristic} found that a valid variability timescale in the light curve should be larger than the mean cadence of the light curve and less than ten percent of the baseline.
We adopted this criterion to find out the unreliable regions in the LSP and shaded areas (grey) in Figure \ref{Fig-LSP_DRW} represent the invalid regions in the LSP.\par

In this work, we report a potential transient quasi-periodic oscillation with a period of approximately 411 days, detected in the $\gamma$-ray emission from the blazar S5 1803+784. The signal exceeds a 3$\sigma$ significance level across multiple analysis methods, supporting its authenticity.

\begin{figure}
    \centering
    \includegraphics[width=0.47\textwidth]{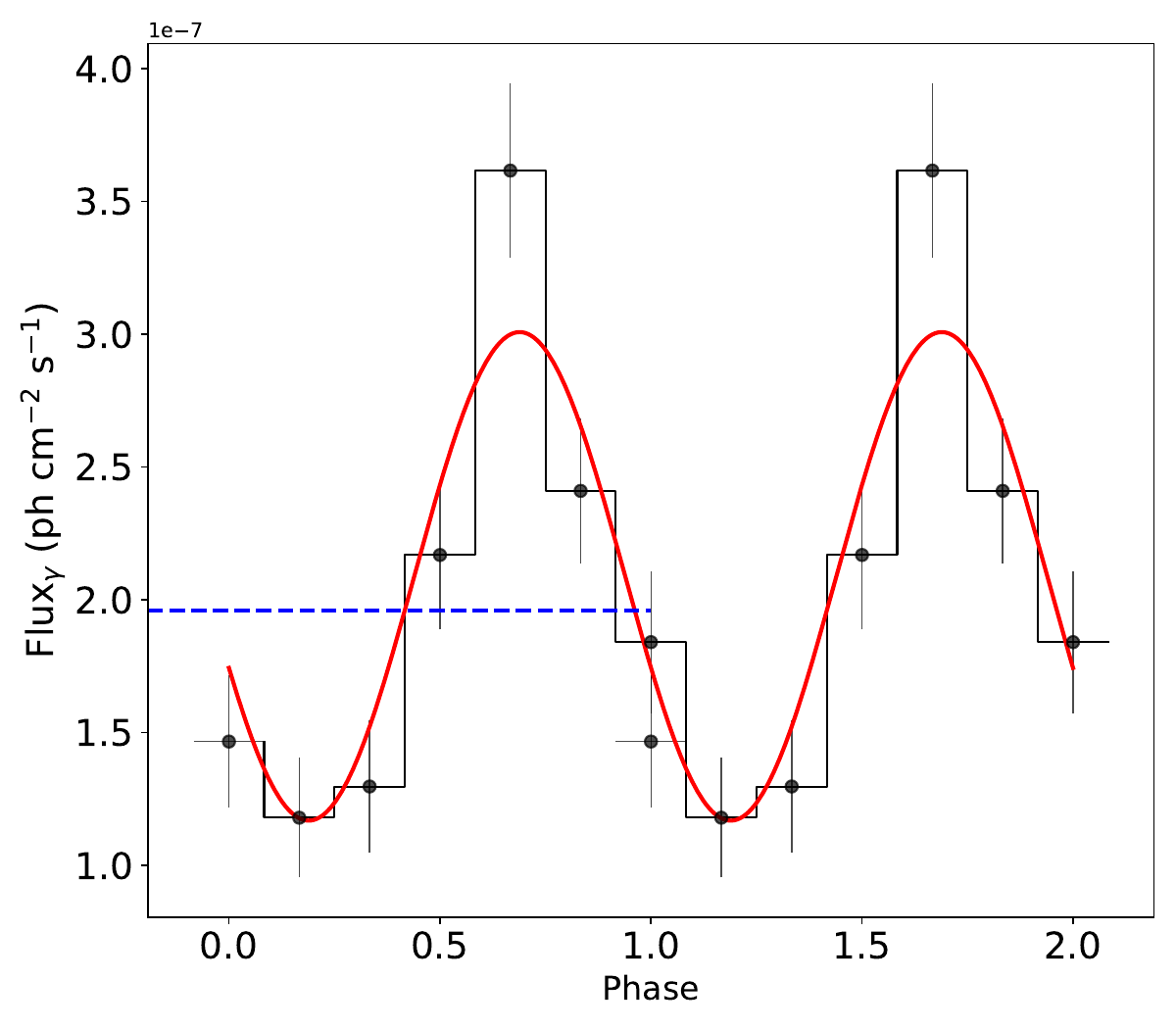}
    \caption{The folded Fermi-LAT light curve (black) of blazar S5\,1803+784 in the time domain from MJD 54683 -- 60187 above 100 MeV with a period of $\sim$411 days. The dashed blue line represents the mean value and best-fitted sine function to data in red. We present two-period cycles here for better clarity.}
    \label{Fig-phase}    
\end{figure}

Further confirmation of the observed periodic signal, we constructed a phase-folded $\gamma$-ray light curve with a period of $\sim$411 days and fitted it with a sine function, see Figure \ref{Fig-phase}. It further confirms a transient periodic signal in the light curve of blazar S5\,1803+784. To enhance the visual clarity, we present two complete periodic cycles within the phase-resolved $\gamma$-ray light curve. 

\begin{figure*}
    \centering
    \includegraphics[width=0.7\textwidth]{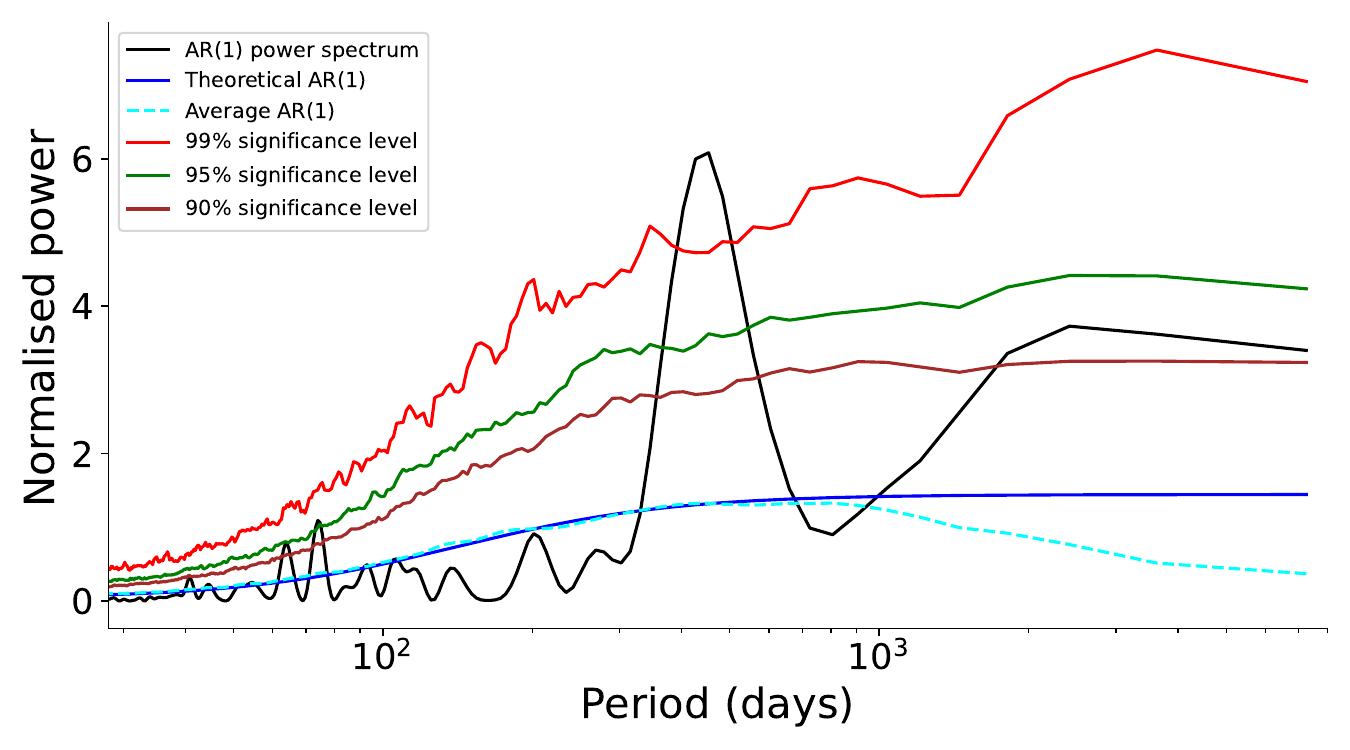}
    \caption{Analysis of the light curve using the AR(1) process with the REDFIT tool. The red noise-corrected power spectrum (black) is presented, alongside theoretical and average AR(1) spectra. The significance levels of 99$\%$, 95$\%$, and 90$\%$ are indicated in red, green, and brown, respectively. A clear transient QPO signature with period of $\sim$421 days is present in the $\gamma$-ray light curve of blazar S5\,1803+784.}
    \label{Fig-redfit}    
\end{figure*}

%###################    DRW model    #######################

%$$$$$$$$$$$$$$$$$$$$$$$$$$$$$$$$$$$$$$$$$$$$$$$$$$$$$$$$$$$$

\begin{table*}
\setlength{\extrarowheight}{7pt}
\setlength{\tabcolsep}{7pt}
\centering
\caption{The transient QPO results obtained using four different approaches are summarized here. Column (1) represents the 4FGL name of the blazar S5 1803+784. Columns (2)–(5) present the transient QPO frequencies derived from the LSP, WWZ, REDFIT, and DCDFT methods, with their uncertainties. The local significance of each QPO is indicated in parentheses next to the corresponding frequency value. }

\begin{tabular}{c c c c c}
\hline
\hline
4FGL Name  & $10^{-3}\times$ \ LSP $\left( day^{-1} \right)$ & $10^{-3}\times$ \ WWZ $\left( day^{-1} \right)$ & $10^{-3}\times$ \ REDFIT $\left( day^{-1} \right)$ & $10^{-3}\times$ \ DCDFT $\left( day^{-1} \right)$\\
(1) & (2) & (3) & (4) & (5) \\
[+2pt]
\hline
4FGL J1800.6+7828 &  2.43$\pm$0.45 ($>99.7\%$) & 2.51$\pm$0.2 ($>99.5\%$) & 
2.37$\pm$0.49 ($>99\%$) & 
2.51$\pm$0.268\\  
[+4pt]
\hline
\end{tabular}

\label{tab:QPO_all}
\end{table*}

\subsection{\textbf{\texorpdfstring{$\gamma$-ray spectral analysis}{gamma-ray spectral analysis}}}

\begin{figure}
\centering
\includegraphics[scale=0.45]{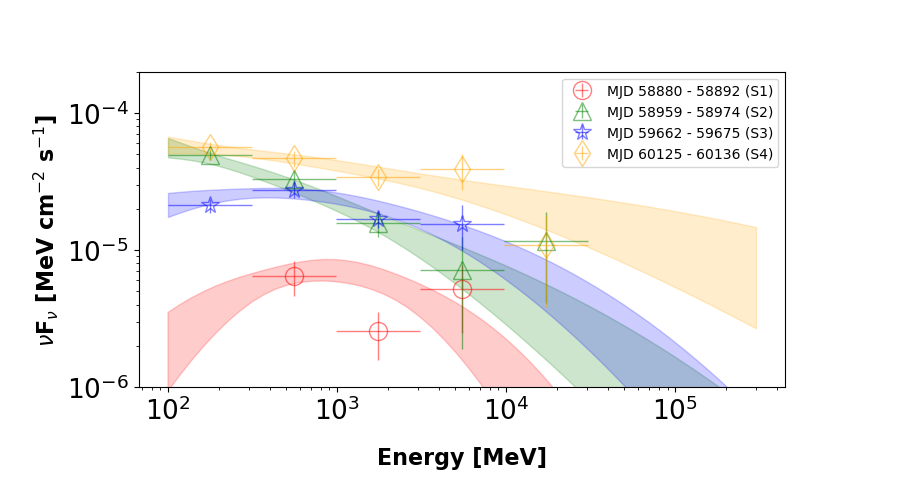}%
\caption{$\gamma$-ray spectral plot using log parabola model}
\label{fig:gp}
    
\end{figure}

%\subsection{$\gamma$-ray spectral analysis}
We performed  $\gamma$-ray spectral analysis using Fermipy's standard procedure. The log-parabola spectral model was applied to our source of interest, as described by\
\begin{equation}
\label{eq:2}
n(E)= K (E/E_{pivot})^{-(\alpha+\beta \log(E/E_{pivot}))}
\end{equation}
where $\alpha$ is the photon-index at pivot energy $E_{pivot}= 1$\,GeV, $\beta$ is the curvature parameter, and K is the normalization.  The $\gamma$-ray spectra of the source were fitted using a log-parabola distribution model. The derived parameters are summarized in Table \ref{table:gamma}, while the corresponding spectral shapes for four distinct epochs are illustrated in Figure \ref{fig:gp}.  The reported spectral form of this particular source is also log parabola in the \textit{4FGL} Catalog \footnote{\url{https://heasarc.gsfc.nasa.gov/W3Browse/fermi/fermilpsc.html}}. Although previous studies by \cite{10.1093/mnras/stac1009} have also shown, that broken power law, and log parabola (LP) are better models to explain $\gamma$-ray SED. Interestingly, we found that states S1 and S2 have large curvature characteristics. Additionally, when the source changed from low to high flux levels, we saw notable shifts in the curvature parameter ($\beta$) and $\alpha_\gamma$ . An important feature of our results is the exceptionally steep  $\alpha_\gamma$  of 2.48 reported between MJD 60125 -- 60150. This value surpasses the previous values of 2.27 and 2.29 reported by \cite{inproceedings, 10.1093/mnras/stac1009} respectively. From the four epochs, the photon spectral index values are \( \alpha_\gamma = 1.65 \) (S1), \( 2.10 \) (S2), \( 2.18 \) (S3), and \( 2.48 \) (S4). These values are consistent with the earlier findings like classification of Changing-Look Blazars (CLBs) into three states: FSRQ (\( \alpha_\gamma \gtrsim 2.2 \)), Transition (\( 2.0 < \alpha_\gamma < 2.2 \)), and BL Lac (\( \alpha_\gamma \lesssim 2.0 \)) (Ren et al., 2024). Specifically:
    S1 (\( \alpha_\gamma = 1.65 \)) falls within the BL Lac state, S2 (\( \alpha_\gamma = 2.10 \)) and S3 (\( \alpha_\gamma = 2.18 \)) lie in the Transition state, S4 (\( \alpha_\gamma = 2.48 \)) corresponds to the FSRQ state. This progression is consistent with Support Vector Machine (SVM) classification, which distinguishes BL Lacs (\( \alpha_\gamma = 2.032 \pm 0.212 \)) from FSRQs (\( \alpha_\gamma = 2.470 \pm 0.20 \)) based on photon spectral index (Fan et al., 2022). Additionally, the density distributions of \( \alpha_\gamma \) for CLBs (2.0–2.5) lie between BL Lacs and FSRQs, supporting their classification as transitional objects (Kang et al., 2024, 2025).

\begin{table*}
\centering  
%\scriptsize
\caption{Details of the best ﬁt parameters obtained by $\gamma$-ray spectrum analysis of S5\,1803+784  using Fermipy tool. In the table, Columns 1: Observations,  2: Particle distribution models, Parameters: 3, 4, 5, 6, 7, 8: represent the physical parameters corresponding to chosen particle distribution.}
 
%\begin{adjustbox}{}

\label{tab:2}
\begin{tabular}{c c c c c c c c c c c c}
\hline
\hline

&&&\\ 
Observation&Model&$\alpha$&$\beta$&Norm&$E_{b}$&TS&loglike \\
(1) & (2) & (3) & (4) & (5) & (6) & (7) & (8)\\
\hline
\\
S1&LP&1.65$\pm$0.41&0.48$\pm$0.33&2.05$\pm$0.52&645.15&69.86&-5774.4496\\
\\
\\
S2&LP&2.10$\pm$0.08&0.10$\pm$0.05&6.13$\pm$0.55&645.15&112.13&-7937.145\\
\\
\\
S3&LP&2.18$\pm$0.06&0.02$\pm$0.04&10.81$\pm$0.90&645.15&885.722&-5509.414\\
\\
\\
S4&LP&2.48$\pm$0.09&0.05$\pm$0.02&6.55$\pm$0.80&645.15&418.494&-5192.92\\
\\
\hline
\label{table:gamma}
\end{tabular}
%\end{adjustbox}
\end{table*}

\subsection{\textbf{Broadband spectral analysis}}
 \label{fig:broadband}
We investigated the broadband spectral features of S5\,1803+784, combining Swift-UVOT/XRT and Fermi data. A one-zone leptonic model, incorporating synchrotron and SSC mechanisms, was used to model the spectral energy distribution and constrain its particle energy distribution. So, we assume emission originates from a spherical region of radius R, with the relativistic jet alignment producing Doppler-boosted emission, accounted for by the Doppler factor, $\delta=\frac{1}{\Gamma(1-\beta\,cos\theta)}$. Relativistic particles within the jet lose energy through synchrotron and IC interactions with magnetic fields and photons. For this, we applied the SSC process, utilizing the jet's own synchrotron photons as seed photons for IC scattering. The electron Lorentz factor, $\gamma$ is expressed in terms of new variable $\xi$, such that $\xi=\gamma\sqrt{\mathbb{C}}$, where $\mathbb{C}=1.36 \times 10^{-11}\frac{\delta B}{1+z}$ with z being the redshift of source.
The synchrotron flux recieved by the observer with $\xi$ as parameter can be written as \citep{10.1093/mnras/stad3534} \\
\begin{equation}\label{eq:syn_flux}
 F_{syn}(\epsilon)=\frac{\delta^3(1+z)}{d_L^2} V  \mathbb{A}  \int_{\xi_{min}}^{\xi_{max}} f(\epsilon/\xi^2)n(\xi)d\xi,
 \end{equation}
 where $\rm d_L$  is luminosity distance, V is the volume of emission region, $\rm \mathbb A = \frac{\sqrt{3}\pi e^3 B}{16m_e c^2 \sqrt{\mathbb{C}}}$,  $\xi_{min}$ and $\xi_{max}$ correspond to the minimum and maximum energy of electron,  and f(x) is the synchrotron emmisivity function \citep{1986rpa..book.....R}. The
 SSC flux received by the observer at energy $\epsilon$  can be obtained using the equation 
 \begin{equation}\label{eq:ssc_flux}
  \begin{split}
 F_{ssc}(\epsilon) =\frac{\delta^3(1+z)}{d_L^2} V  \mathbb{B} \epsilon & \int_{\xi_{min}}^{\xi_{max}} \frac{1}{\xi^2}  \int_{x_1}^{x_2}   \frac{I_{syn}(\epsilon_i)}{\epsilon_i^2}  \\
&  f(\epsilon_i, \epsilon, \xi/\sqrt{\mathbb{C}}) d\epsilon_i   n(\xi)d\xi
 \end{split}
 \end{equation}
where, $\rm \epsilon_i$ is incident photon energy,   $\rm \mathbb{B} = \frac{3}{4}\sigma_T\sqrt{\mathbb{C}}$,  $\rm I_{syn}(\epsilon_i)$ is the synchrotron intensity,  $\rm x_1=\frac{\mathbb{C} \, \epsilon}{4\xi^2(1-\sqrt{\mathbb{C}} \,\epsilon/\xi m_ec^2)}$,  $\rm x_2=\frac{\epsilon}{(1-\sqrt{\mathbb{C}}\,\epsilon/\xi m_e c^2)}$ and

\begin{equation}
f(\epsilon_i, \epsilon, \xi)= 2q\log q+ (1+2q)(1-q)+\frac{\kappa^2q^2(1-q)}{2(1+\kappa q)} \nonumber
\end{equation}
here $\rm q=\frac{\mathbb{C}\epsilon}{4\xi^2\epsilon_i(1-\sqrt{\mathbb{C}}\epsilon/\xi m_ec^2)}$ and $\rm \kappa=\frac{4\xi\epsilon_i}{\sqrt{\mathbb{C}} m_e c^2}$.\\

We numerically solve Equations  \ref{eq:syn_flux} and   \ref{eq:ssc_flux}, and integrate the resulting code as a local convolution model in XSPEC. In the convolved  XSPEC model `energy' variable is interpreted as $\xi = \gamma \sqrt{\mathbb{C}}$. This convolution model allows for statistical fitting of broadband SEDs, making it possible to model broadband spectra for any given particle energy distribution $n(\xi)$.  Within this region, we consider a relativistic particle distribution denoted as n(E). However, Broken power law (BPL) distribution provides reasonable parameters for the broadband SED modeling. The BPL distribution is given by
\begin{equation}
\label{eq:3}
n(E) = 
\begin{cases}
    K E^{-p_1} & \text{for } E < E_b \\
    K E_b^{p_2-p_1} E^{-p_2} & \text{for } E \geq E_b
\end{cases}
\end{equation}
where $K$ is normalization, $E_{\rm b}$ is break energy, $p_{1}$ and $p_{2}$ indicate the particle indices before and after break energy $E_\mathrm{b}$.\par
In the broadband spectral energy distribution (SED) modeling, key parameters such as the bulk Lorentz factor ($\Gamma$), magnetic field strength (B), size of the emission region (R), and jet opening angle ($\theta$) play a crucial role. In the context of BPL model, which is characterized by four parameters like spectral indices before and after the break ($p_{1}$ and $p_{2}$),  the break energy $E_\mathrm{b}$, and the normalization factor (Norm), the normalization factor is typically treated as a free parameter during the fitting process. Initial values for these parameters are chosen based on the observed spectral shapes and flux levels of the synchrotron and SSC components. These parameters are then iteratively varied to converge on their optimal values, ensuring a good fit to the data. The broadband SED fitting code also allows for the inclusion of the jet power ($P_{jet}$) as one of the parameters. 
 Adopting the standard approach, where cold protons and non-thermal electrons equally contribute to jet inertia, we derived an estimate of the total jet power formula by \cite{ Ghisellini_2014}
\begin{equation}
P_{\text{jet}} = \pi R^2 \Gamma^2 \beta c U_e
\end{equation}
where 
We denote $\Gamma$ as the bulk Lorentz factor, $\beta  = \frac{v}{c}$  as the jet velocity (v/c), R as the jet radius, and $U_e$ as the energy densities of electrons respectively.

\begin{figure}
\centering
\includegraphics[scale=0.37]{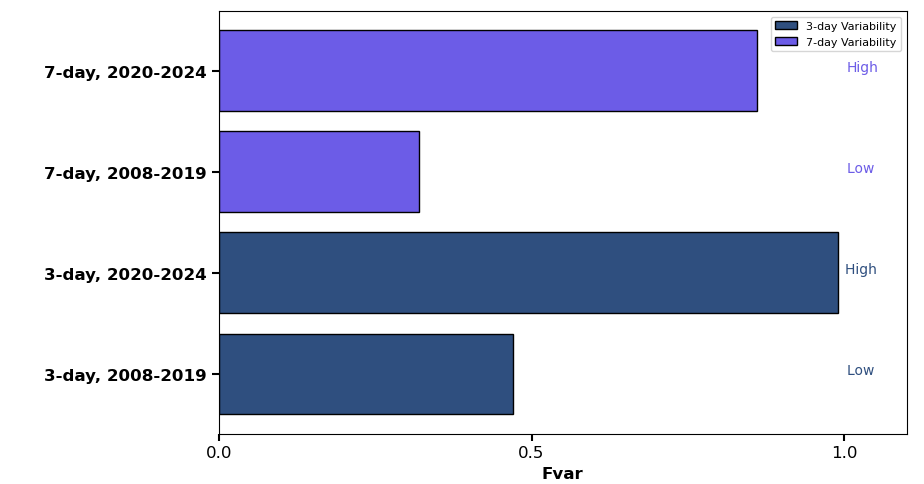}%

\caption{Fractional variability $(F_{var})$ in different flux states and time bins for $\gamma$-ray data. The $F_{var}$ values are calculated for various flux states, including the low-flux (LFS) and high-flux (HFS). The time bins considered are  3-day and 7-day bins.}
\label{fig:fv_plot}

\end{figure}

\begin{table*}
\centering
%\scriptsize
\caption{The table presents the optimal parameters obtained by fitting the broadband spectrum of S5 1803+784 using a one-zone SED model in XSPEC. The columns represent: (1) Observation details, (2) Particle distribution models used, (3) Bulk Lorentz factor, (4-6) Parameters corresponding to the chosen particle distribution, (7) Magnetic field strength (in Gauss), (8) Logarithmic jet power (in erg $sec^{-1}$), and (9) Reduced-$\chi^2$. Subscript and superscript values indicate lower and upper bound errors, respectively. A dash (--) indicates that the lower or upper bound error is unconstrained.}
 
%\begin{adjustbox}{}

\label{tab:3}
\begin{tabular}{c c c c c c c c c c c c }
\hline
\hline

&&&\\ 
Observation&Model&$\Gamma$&$p_{1}$&$p_{2}$&$ \text{log R} \text{(cm)}$&B(G)&$P_{jet}$&$\chi^2_{red}$\\
(1) & (2) & (3) & (4) & (5) & (6) & (7) & (8)&(9)\\
\hline
\\
S1&BPL&$14.90^{+0.48}_{-0.37}$&$1.24^{+0.12}_{-0.20}$&$4.34^{+0.34}_{-0.22}$&16.0&$0.26^{+0.03}_{-0.02}$&$43.40^{+0.20}_{-0.17}$&0.82\\
\\
\\
S2&BPL&$30.36^{-}_{-0.60}$&$2.18^{+0.06}_{-0.06}$&$5.27^{+0.85}_{-0.21}$&16.0&$0.26^{+0.04}_{-0.03}$&$46.03^{-}_{-0.80}$&1.16\\
\\
\\
S3&BPL&$22.56^{+0.52}_{-0.43}$&$1.97^{+0.02}_{-0.03}$&$4.99^{+0.37}_{-0.29}$&16.0&$0.05^{+0.003}_{-0.002}$&$45.85^{+0.25}_{-0.29}$&1.09\\
\\
\\
S4&BPL&$21.52^{+0.22}_{-0.17}$&$1.10^{+0.05}_{-0.05}$&$4.85^{+0.15}_{-0.12}$&16.0&$0.19^{+0.02}_{-0.01}$&$45.36^{+0.16}_{-0.13}$&0.56\\
\\
\hline
\label{table:3}
\end{tabular}
%\end{adjustbox}
\end{table*}

 \begin{figure*}
\centering

\includegraphics[scale=0.32,angle=0]{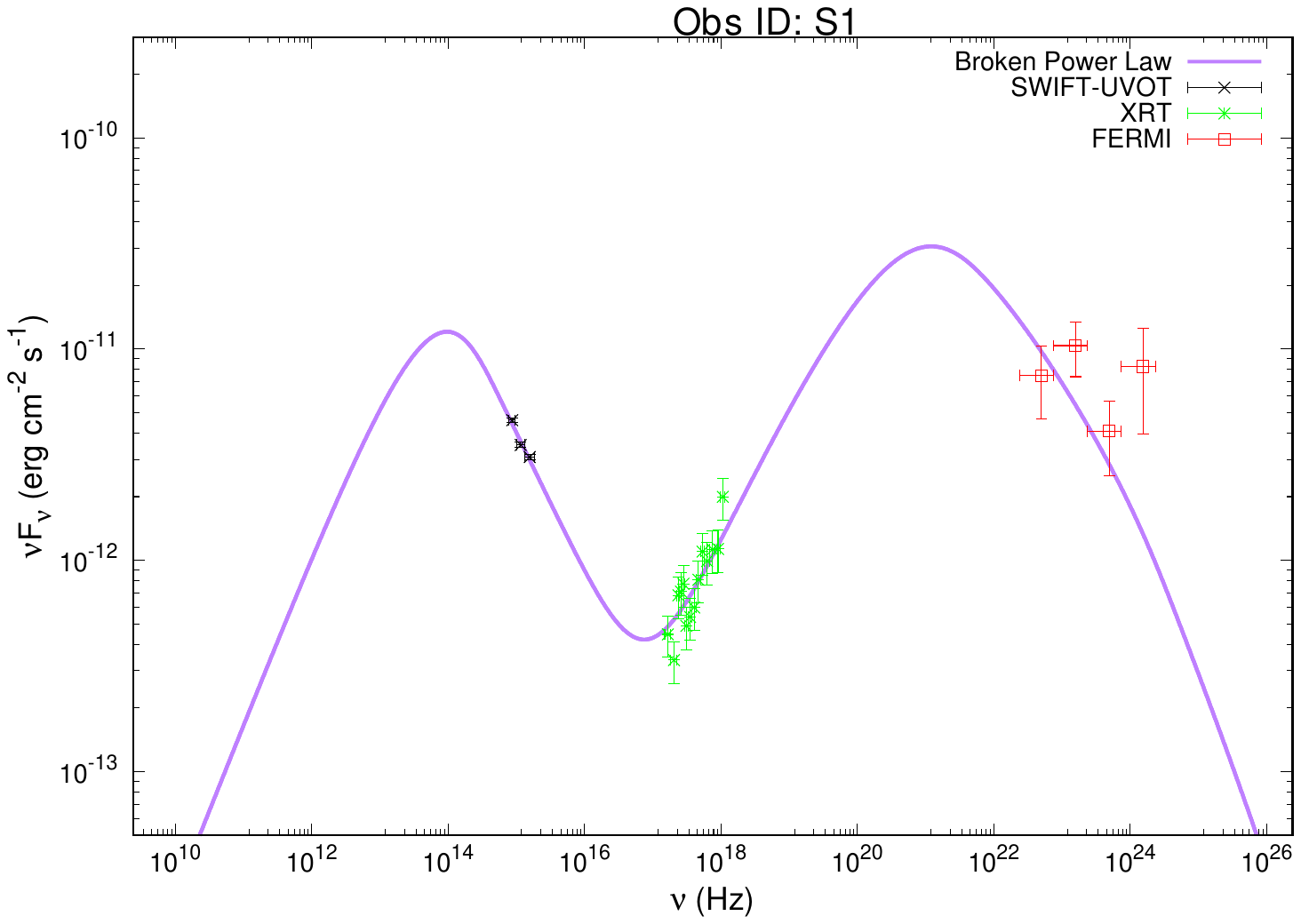}%
\includegraphics[scale=0.32,angle=0]{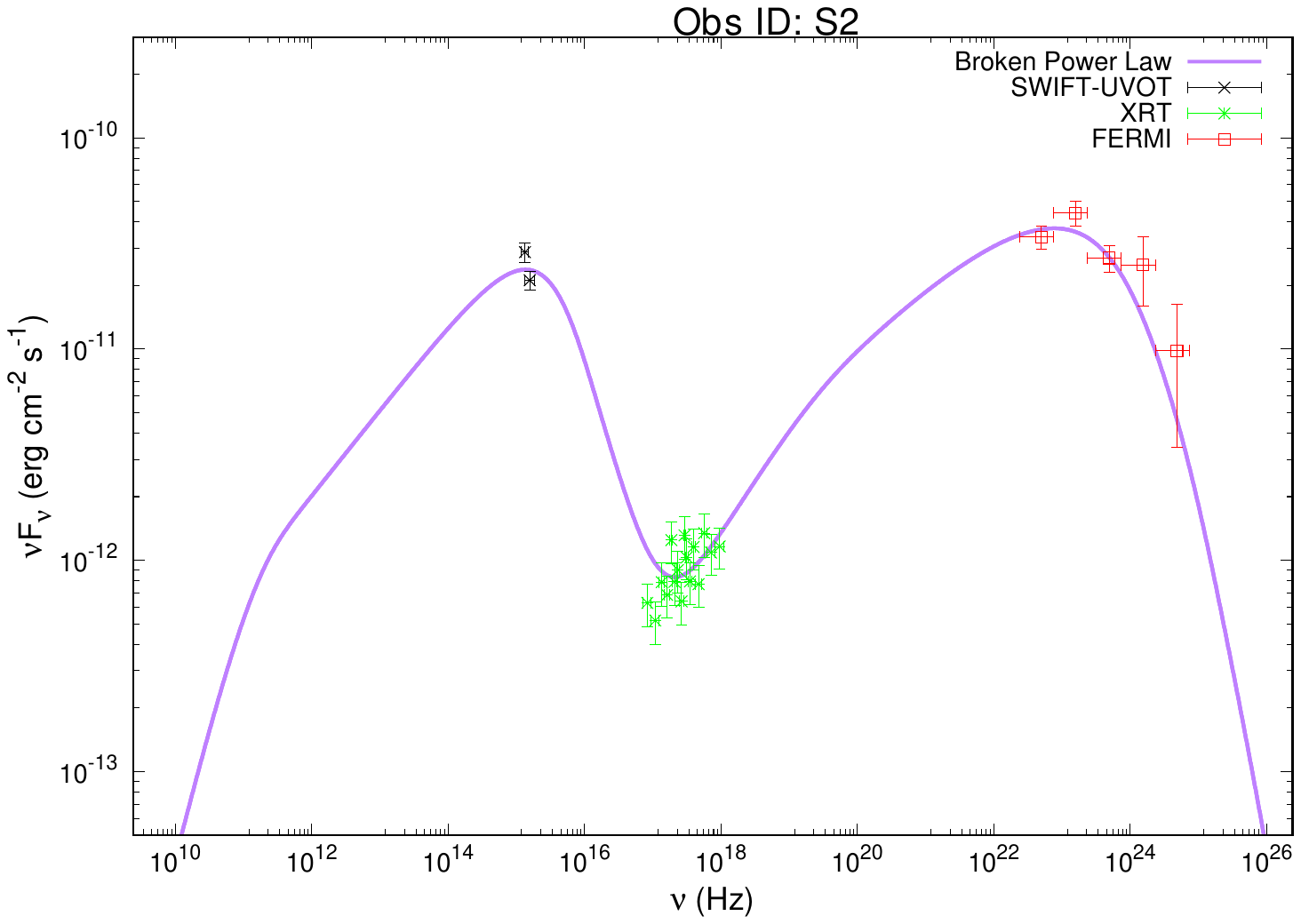}\\%
\includegraphics[scale=0.32,angle=0]{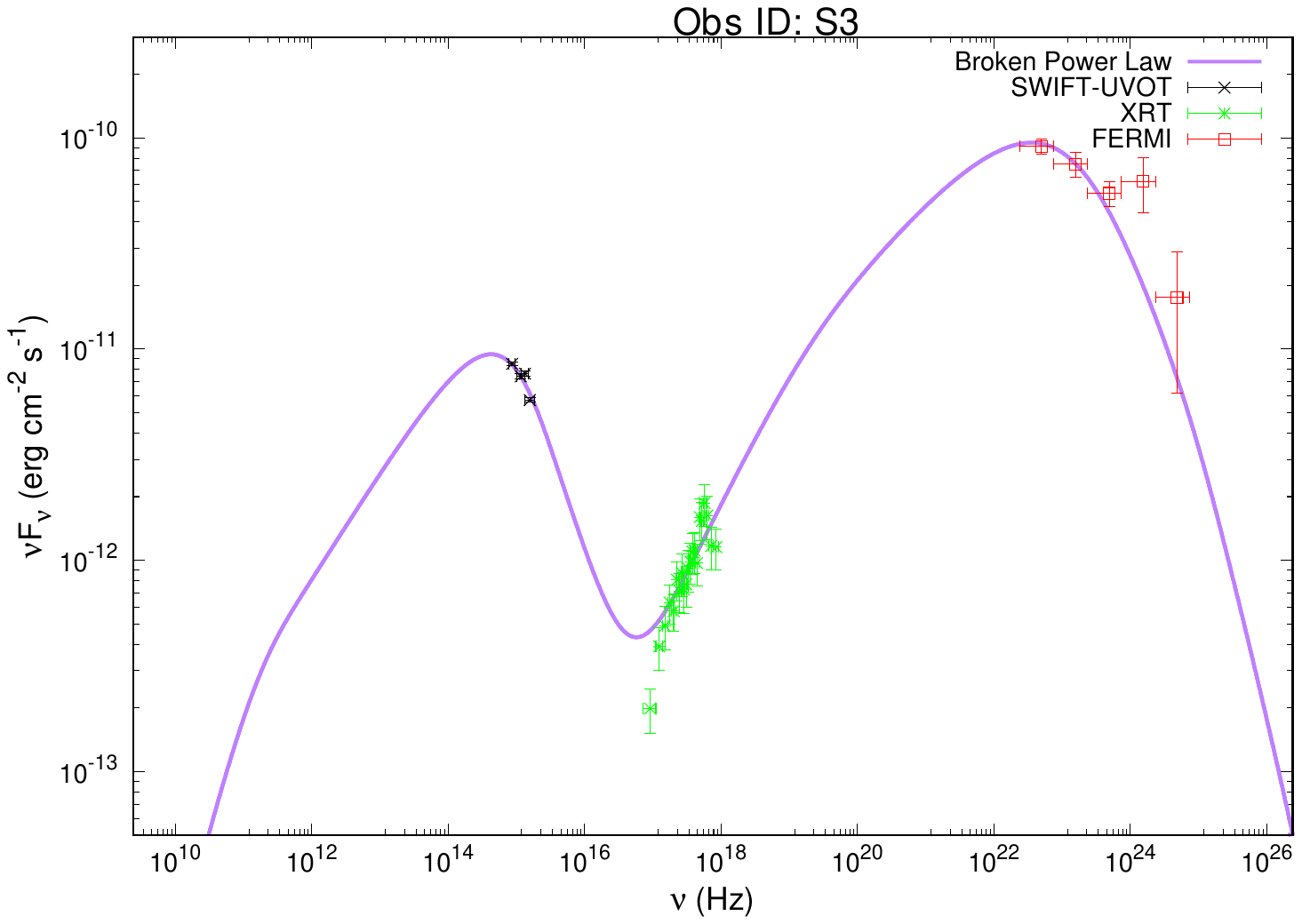}%
\includegraphics[scale=0.32,angle=0]{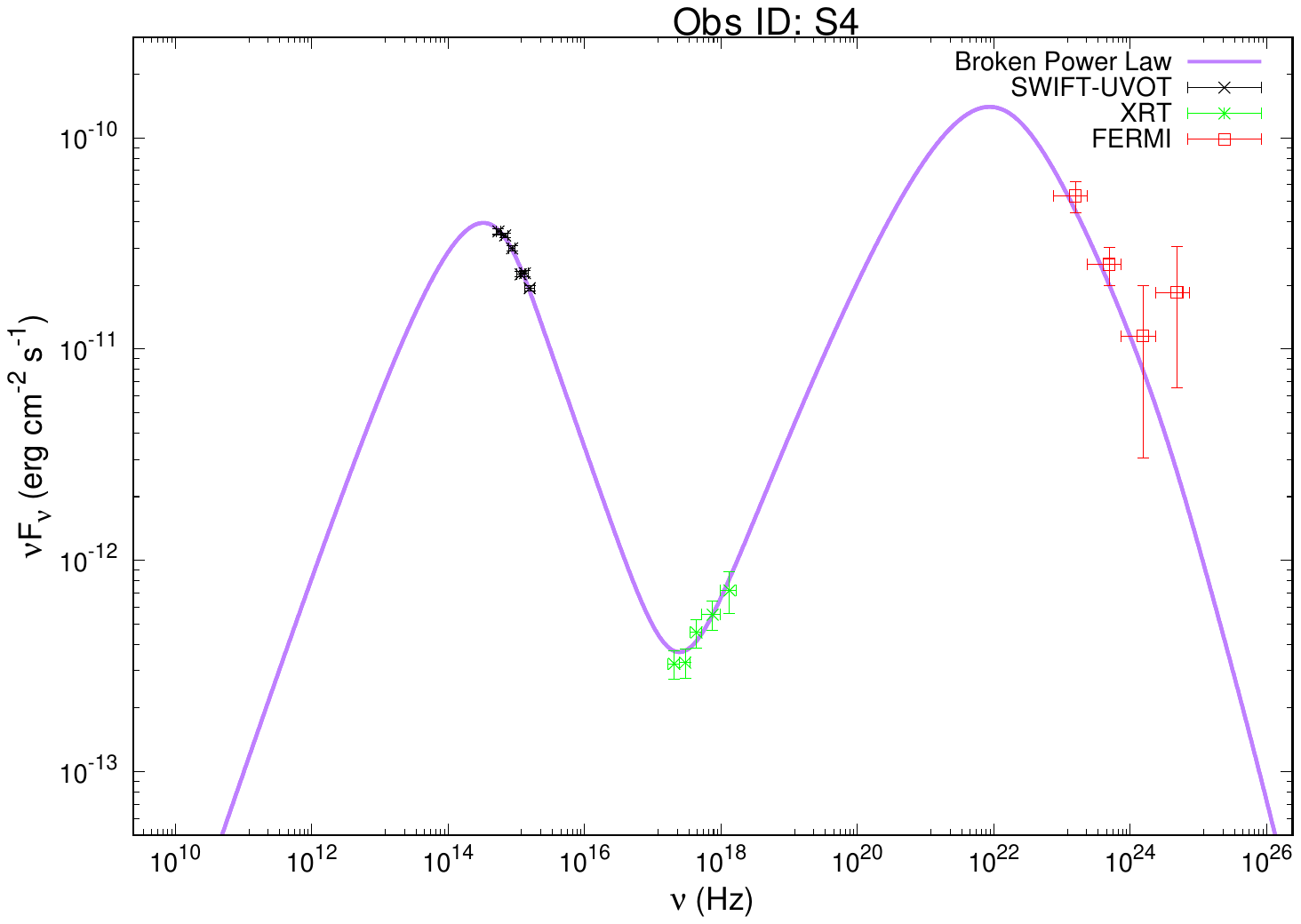}

\caption{ The following plots show the broadband spectral energy distributions (SEDs) for four distinct epochs (S1-S4) using the Broken Power Law (BPL) particle distribution model} 
%Top left panel is a variation of $\rm P_{jet}$ with $\Gamma$, top right panel is a plot of jet power with $\gamma_{min}$, the bottom left panel is a plot of $\rm P_{jet}$ with R. Bottom right panel represents the variation of R with a B.}
%\label{fig:cor_log}
\end{figure*}

%\begin{figure*}
%\centering
%\includegraphics[scale=0.5]{bkn1.eps}
%\includegraphics[scale=0.5]{bkn2.eps}\\
%\includegraphics[scale=0.5]{bkn.eps}
%\includegraphics[scale=0.5]{size_bkn.eps}\\
%\caption{ Jet power variation with the bulk Lorentz factor, minimum particle energy and emission region size using the BPL model for the particle distribution. The top left panel is a variation of $\rm P_{jet}$ with $\Gamma$, top right panel is a plot of jet power with $\gamma_{min}$, the bottom left panel is a plot of $\rm P_{jet}$ with R.   Bottom right panel represents the variation of R with a B.}
%\label{fig:cor_bkn}

%\end{figure*}

 \begin{figure}
    \centering
    \includegraphics[width=0.5\textwidth]{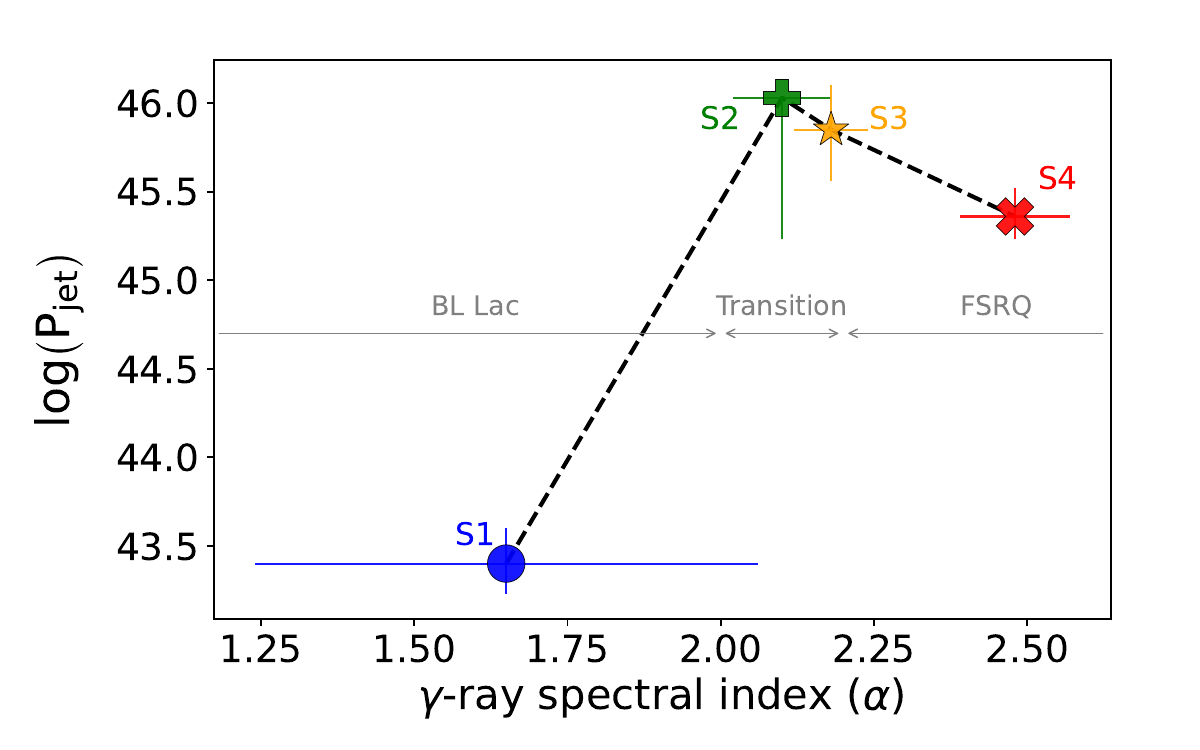}
    \caption{Jet power ($\log P_{\rm jet}$) as a function of $\gamma$-ray spectral index ($\alpha$) across four observational epochs: S1, S2, S3, and S4. The photon spectral index starts at 1.65 in S1 and progressively increases through subsequent epochs: 2.10 in S2, 2.18 in S3, and 2.48 in S4. This trend suggests that the blazar S5\,1803+784 may be undergoing a transition from a BL Lac to FSRQ. }
    \label{Fig-Jet_power_slope}    
\end{figure}

\section{Summary and Discussion}
\label{sum}
The transition blazar S5\,1803+784 offers a unique opportunity to explore blazar physics due to its remarkable ability to shift between various activity phases. This characteristic has made it a continuous subject of study, with several multi-wavelength investigations conducted during periods of flaring activity. However, simultaneous X-ray/UV/Optical data have been limited \cite{10.1093/mnras/stac1009, inproceedings}, highlighting the need for further analysis. This study offers a detailed temporal and spectral analysis of S5\,1803+784, exploring both its quiet and flaring states, and uncovering its underlying properties.  Notably, we observed the highest $\gamma$-ray flux of ($\mathrm{2.26\pm0.062)\times10^{-6}~phcm^{-2}s^{-1}}$  MJD 59667.29 -- 59676.03. Temporal analysis of $\gamma$-ray data revealed striking correlations between activity levels and flux fluctuations.  Variability is crucial in understanding blazar structure, as it can arise from intrinsic factors. Intrinsic mechanisms involve fluctuations in particle injection or acceleration within specific jet regions, such as the base or shock front. These fluctuations are amplified by shock interactions and magnetic field changes, leading to pronounced observed flux variations due to Doppler boosting. Our analysis revealed significant variability factors in both 3-day and 7-day binned light curves. Specifically, the three-day binned light curve displayed a variability factor of 0.47 ± 0.03 between 2008-2019, increasing to 0.99 ± 0.03 during 2020-2024. Similarly, the seven-day binned light curve showed variability factors of 0.32 ± 0.06 and 0.86 ± 0.05 for the same periods as shown in Figure \ref{fig:fv_plot}. Surprisingly, X-ray emissions during the period (2020-2023) were lower than UV/optical and $\gamma$-ray wavelengths, diverging from  BL\,Lac behavior. As high-energy electrons characterized by rapid cooling, generate substantial variability in the Optical/UV and $\gamma$-ray domains. In contrast, the more gradual cooling of lower-energy electrons leads to reduced variability in the X-ray domain. But in the case of BL\,Lacs,  X-ray fluctuations show greater amplitude than those in $\gamma$-ray and optical regimes, even when the source is in a low-flux state \cite{Kapanadze_2023, abe2023multi, Abe:2024eba}. The unusual behavior hints at Low-energy-peaked BL\,Lac (LSP) characteristics, resembling Flat Spectrum Radio Quasar (FSRQ) properties. The lower variability observed in the X-ray band compared to the optical and gamma-ray bands suggests characteristics of a Low-energy-peaked BL Lac (LSP). This behavior can be explained by the shape of the SED in LBLs, where the X-ray emission lies near the lower energy tail of the synchrotron self-Compton (SSC) component. In LBLs, the X-ray emission is primarily produced by lower energy electrons in the jet. These electrons have longer cooling timescales compared to higher energy electrons, as cooling time is directly proportional to electron energy. Higher energy electrons, responsible for optical and gamma-ray emission, lose energy more rapidly through synchrotron and inverse Compton processes, leading to faster variability in these bands. In contrast, the slower cooling of lower energy electrons results in more stable and less variable X-ray emission. Recent studies on the transition blazar B2 1308+326 by \cite{2024A&A...681A.116P} reveal lower X-ray emission compared to optical and $\gamma$-ray bands, aligning with our findings and demonstrating a similar trend in X-ray emission behavior. Examining the structure of fractional variability across different energy levels can provide significant insights into the particle population or processes generating broadband emissions. This study highlights the importance of continued multi-wavelength monitoring of S5\,1803+784 to understand its unique characteristics further to understand the underlying physics of transition blazars.
\par
In addition, A transient quasi-periodic oscillation with a period of $\sim$411 days was identified using multiple methods, including the Lomb-Scargle Periodogram, Weighted Wavelet Z-Transform, and REDFIT analysis. The results from all approaches are consistent and mutually supportive within their error bars. The detection significance of the periodic signals exceeded 99$\%$ across all methods, affirming the authenticity of the signal. \par
%A plausible driving mechanism for this phenomenon is the helical motion of a magnetized plasma blob within the relativistic jet.} \par  
The production mechanism of quasi-periodic oscillations in AGN remains a subject of debate. However, several intriguing physical models have been proposed to explain the observed phenomena. Among these, the supermassive binary black hole (SMBBH) system has been notably successful in accounting for the $\sim$12 yr QPO observed in OJ 287. This model has since been widely adopted to explain other stable, long-term QPOs \cite{valtonen2008massive, villforth2010variability}. In addition to the SMBBH model, other models such as continuous jet precession and Lense-Thirring precession of the accretion disk have been proposed to explain long-duration QPOs \cite{romero2000beaming, rieger2004geometrical, stella1997lense, liska2018formation, you2018x}. But for transient QPOs, which persist for a few cycles, alternative physical models need to be considered. One possibility involves hotspots orbiting within the innermost stable circular orbit (ISCO) of an SMBH. The motion of the hotspots can modulate the seed photon field for the external inverse Compton scattering (EC) within the jet, resulting in periodic gamma-ray emission \cite{zhang1990rotation, gupta2008periodic}. Another potential mechanism involves magnetic reconnection events, where periodically spaced magnetic islands within the jet drive the QPOs of the BH systems at various scales \cite{huang2013magnetic}. It is necessary to explore other mechanisms that might be more favorable for such observed transient quasi-periodic signatures.\par
The helical motion of an enhanced emission region (or blob) within the jet has been extensively studied in the context of transient QPOs during flares, e.g. \cite{mohan2015kinematics, sobacchi2016model, zhou201834} In the one-zone leptonic model, blazar emission primarily originates from this emission region (or blob) in the jet. High-energy gamma-ray emission is produced via the synchrotron self-Compton (SSC) process. A helical jet structure can arise due to large-scale magnetic fields or hydrodynamic effects that induce helical motion in the jet or its emission region (or blob) \cite{mohan2015kinematics, sobacchi2016model, sarkar2021multiwaveband}. Periodic fluctuations in the observed emission are attributed to variations in the Doppler boosting factor as the viewing angle of the moving plasma blob changes. The variability timescale, influenced by factors such as the Doppler boosting factor, pitch angle, and viewing angle, can span from a few days to several months. \par
In this model, the periodic changes in the viewing angle ($\theta$) are driven by the helical motion of the emission region, defined as

\begin{equation}
    \rm{cos} \ \theta_{obs}(t)=\rm{sin\phi} \ \rm{sin\psi} \ \rm{cos (2\pi t/P_{obs})} + \rm{cos\phi} \ \rm{cos\psi}
\end{equation}

where P$_{obs}$ is the observed periodicity, $\phi$ is the pitch angle of the blob, and $\psi$ is the viewing angle or inclination angle measured between the line of sight of the observer and the jet axis. The varying Doppler factor is defined as $\delta = \frac{1}{\Gamma \left[ 1 - \beta \rm{cos \theta_{obs} (t)} \right]}$, where $\Gamma = 1/\sqrt{1 - \beta^2}$ represents the bulk Lorentz factor associated with the motion of blob, and $\beta = \it{v_{jet}}/c$. In the case BL Lac, considered values of $\phi = 2^{\circ}$, $\psi = 5^{\circ}$, and $\Gamma = 8.5$, \cite{abdo2010spectral, sobacchi2016model, zhou201834, sarkar2021multiwaveband, sharma2024detection, chen2024transient}. The relation between the observed and rest-frame periods is defined as 

\begin{equation}
    P_{rest} = \frac{P_{obs}}{1 - \beta \ \rm{cos\phi} \ \rm{cos\psi}}
\end{equation}

\begin{figure}
    \centering
    \includegraphics[width=0.48\textwidth]{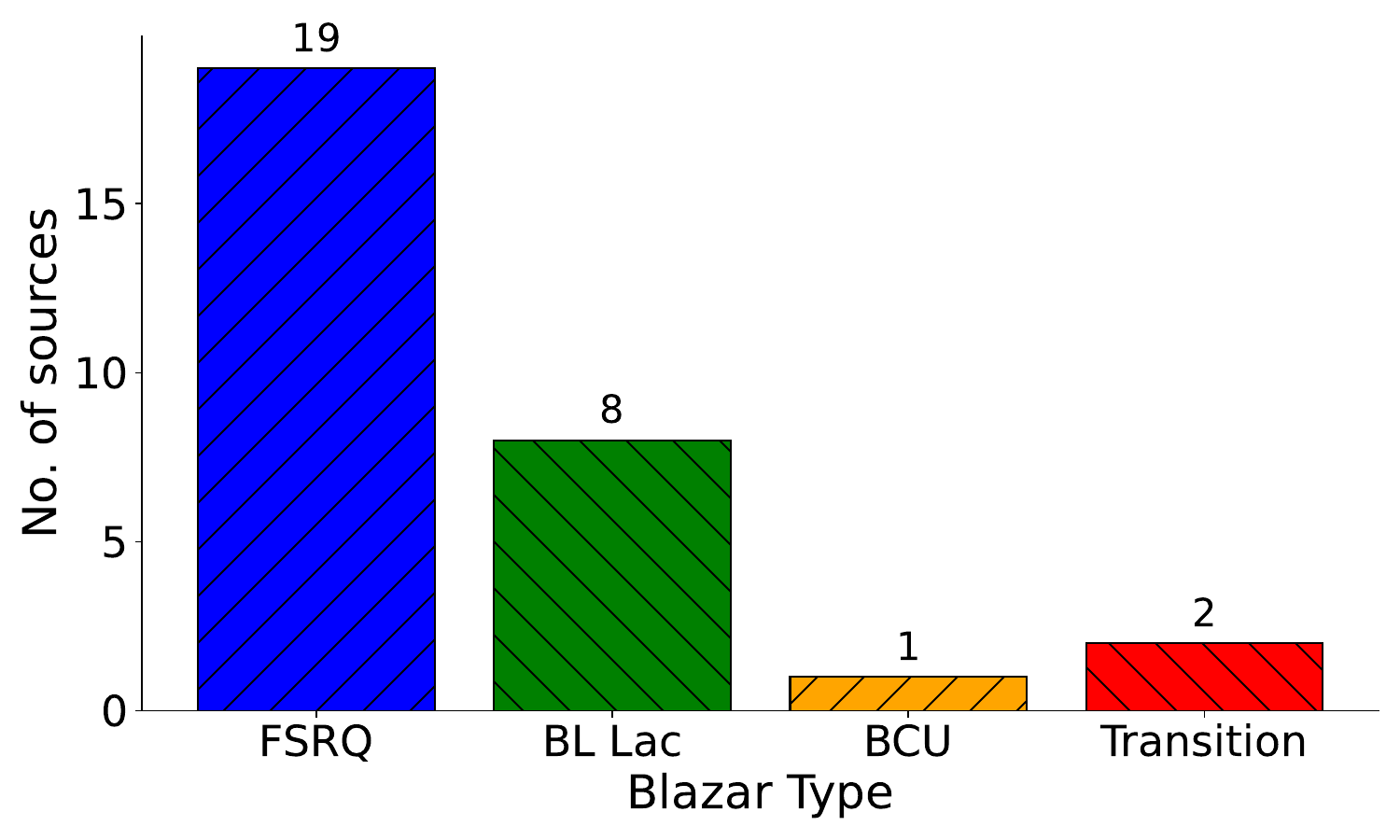}
    \caption{Distribution of reported transient quasi-periodic oscillations (QPOs) observed in different classes of blazar in the literature Table \ref{tab: Transient_QPO_2}, with timescales ranging from days to years. The figure categorizes sources based on their type, including Flat Spectrum Radio Quasars (FSRQs), BL\,Lacs, Blazar Candidates of Uncertain Type (BCUs), and transition blazar from BL Lac to FSRQ.}
    \label{Fig-Distribution}    
\end{figure}

The estimated $\beta$ value using the expression of $\Gamma$ is 0.99305. The translated period value in the rest frame of the blob is $P_{rest} \sim$ 99.3 yr for the $P_{obs}$=411 days and the distance traveled by the blob in one cycle is $D_{1P} = c \ \beta \ P_{rest} \ cos\phi \approx 30.18 pc$ and the projected distance for entire transient phase is $D_{projected}=N \ D_{1P} \ sin\psi \approx 7.87 pc$. The extended jets with parsec-scales have been identified in numerous blazars \cite{bahcall1995hubble, vicente1996monitoring, tateyama1998observations}, and the study is supported by the optical polarization observations highlighting the evidence of the presence of helical structures \cite{marscher2008}. However, the exact origin of their helical structure remains unclear. Further, a modified helical jet model might be a more robust mechanism to explain such features, in which a blob moves helically inside a time-varying jet. In this model, the inclination angle between the jet axis and the line of sight could be time dependent, $\psi \equiv \psi(t)$ \cite{sarkar2021multiwaveband, prince2023quasi}. The variation in the angle could be attributed to the periodic modulations in observed emissions and outliers (high flux points) within the light curve could be associated with some stochasticity present in the jets. In this investigation, the conclusion is that the helical magnetic field could play a key role in shaping these jet structures. \cite{chen2024transient} noticed that the transient QPOs are more commonly observed in FSRQ than BL Lac. A distribution of sources belonging to different blazar classes reported as a transient QPO source is given in Figure \ref{Fig-Distribution} and detailed in Table \ref{tab: Transient_QPO_1}, \ref{tab: Transient_QPO_2}.\par

The Fermi-LAT detection of various blazar types, including Low-Synchrotron-Peaked (LSP), Intermediate-Synchrotron-Peaked (ISP), and High-Synchrotron-Peaked (HSP) blazars, has significantly advanced our understanding of these enigmatic objects \cite{ackermann2015third}. However, a substantial fraction of LSP blazars remain undetected, suggesting that they may possess distinct properties that hinder their detection. In fact, research suggests that these undetected LSP blazars may constitute a distinct class of “$\gamma$-ray quiet blazars" characterized by lower polarization and notable radio brightness, as evidenced by optical polarization properties and radio emission \cite{2016MNRAS.463.3365A}.
This phenomenon is attributed to the steep $\gamma$-ray photon spectral index of LSP blazars, which poses a significant detection challenge  \cite{ackermann2015third, Paliya_2020}. Furthermore, this effect is exacerbated for high-redshift sources, where the $\gamma$-ray spectral energy distribution (SED) is shifted to lower frequencies in the observer's rest frame. In addition, the study suggests that the lower synchrotron peak blazars (LSPs) could be the source of changing-look blazars (CLB)(BL\,Lac to FSRQ) because 90.7 percent of the  CLB are found to be LSP \cite{Xiao_2022, 10.1093/mnras/stad2456}.  Recently, a study by \citet{2024A&A...685A.140R}, investigated the evolution of Changing-Look Blazars (CLBs) on long timescales, classifying them into three morphological types: FSRQ, Transition, and BL\,Lac states, based on the relationship between emission line equivalent width and $\alpha_{\gamma}$ photon spectral index, revealing distinct $\alpha_{\gamma}$ ranges for each state found that $\alpha_{\gamma}$ satisfy the relationships with $\alpha_{\gamma} $ $\gtrsim$ 2.2 for the FSRQ state,  $2.0 < \alpha_{\gamma}  <  2.2$ for the transition state, and $\alpha_{\gamma}$ $\lesssim$ 2.0 for the BL\,Lac state. Alternatively, the $\gamma$-ray spectrum of the source follows a power-law distribution:
$N_\nu = A\nu^{-a}$
with $\alpha_{\gamma}$  = 2.26 ± 0.02, indicating a transitional state between  FSRQs and LSP  (BL\,Lacs), detected with Fermi-LAT observations \cite{Ajello_2020, 10.1093/mnras/stab501}.
Earlier, a study has found that $\gamma$-ray sources detected by Fermi-LAT tend to have exceptionally high jet speeds, corresponding to large bulk Lorentz factors \cite{Lister_2009, 2015ApJ...810L...9L}. It also suggested that LSP blazars with low Lorentz factors and synchrotron peaks below $10^{13.4}$ Hz are less likely to be $\gamma$-ray detectable. Furthermore, a comprehensive analysis of LSP blazars was performed to determine key characteristics, including Compton dominance, peak powers of synchrotron and inverse Compton radiation, and frequency distributions\cite{2018A&A...616A..20A}. In our work,
the $\gamma$-ray spectral analysis of S5\,1803+784 shows  $\alpha_{\gamma}$ = 2.48 and  Compton dominance (CD), indicating a characteristic property of FSRQ sources. Alternatively, a steep $\gamma$-ray slope indicates that the decrease in radiative cooling allowed relativistic particles to accelerate to higher energies, resulting in a corresponding shift in the peaks of the synchrotron and IC components. Also, the $\gamma$-ray spectra of FSRQs are generally softer, and their variability is stronger than those of BL\,Lacs, confirming earlier studies by \cite{Ajello_2020}. The jet power against the $\alpha_{\gamma}$ is given in Figure \ref{Fig-Jet_power_slope}, a remarkable evolutionary trend is uncovered in the blazar S5\,1803+784, as its jet power correlates with a steady increase in $\alpha_{\gamma}$  across four observational epochs (S1-S4). From an initial index of 1.65 in S1, the photon spectral index rises progressively to 2.10, 2.18, and ultimately 2.48 in S4. This sustained shift implies a profound change in the blazar properties, possibly indicating a transition from a BL Lac object to an FSRQ. The jet power values and  $\alpha_{\gamma}$ are given in Table \ref{table:gamma},\ref{tab:3}.
 Recent work employed Support Vector Machine (SVM) classification to distinguish between BL Lac objects and Flat Spectrum Radio Quasars (FSRQs) based on photon spectral index and photon flux \citep{2022Univ....8..436F}.
The average  $\alpha_{\gamma}$ = 2.470 ± 0.201 for 795 FSRQs and $\alpha_{\gamma}$ = 2.032 ± 0.212 for 1432 BL Lacs. Contemporary, \cite{Kang_2024, Kang_2025}, the density distributions of the  $\alpha_{\gamma}$ were analyzed for a sample of 2250 blazars, including 1397 BL Lacs, 105 CLBs (classified as intermediate blazars), and 748 FSRQs using the  Gaussian mixture modeling clustering algorithm. The results revealed that the density distributions of CLBs in terms of $\alpha_{\gamma}$ are distinctly positioned between those of BL Lacs and FSRQs. Specifically, the values of  $\alpha_{\gamma}$ for CLBs were found to lie predominantly in the range of 2.0 to 2.5, indicating an intermediate spectral behavior between the harder spectra of BL Lacs and the softer spectra of FSRQs. This finding supports the classification of CLBs as transitional objects bridging the gap between BL Lacs and FSRQs in terms of their spectral properties.These results collectively emphasize the importance of considering the photon spectral index and other properties to understand blazar diversity.\par

 We applied a leptonic one-zone model to the broadband spectral energy distribution (SED) of S5 1803+784, incorporating both synchrotron and Synchrotron Self-Compton (SSC) effects. Notably, our broadband SED modeling reveals that the SSC model accurately reproduces the observed flux states within this framework, without requiring external Compton contributions. This approach offers several advantages, including seamless integration with XSPEC, enabling reliable and precise statistical analysis, and tightly constrained model parameters \cite{Tantry:2024qvs}. Consequently, multiband observations of blazars facilitate a comprehensive understanding of their flux and spectral evolution. These observations provide critical insights into the spatial structure, dynamics, and electron cooling processes within relativistic jets, thereby enabling rigorous testing of theoretical frameworks. Notably, previous broadband SED analyses revealed that S5\,1803+784 shifted from a low-synchrotron-peaked (LSP) state to an intermediate-synchrotron-peaked (ISP) state during a flare, suggesting the presence of highly energetic particles or a decrease in Compton dominance (CD) \cite{inproceedings}. This transition caused X-ray emission to shift from inverse Compton scattering to the high-energy tail of synchrotron emission. Alternatively, for another transition blazar B2 1308+326, the same trend is followed by X-ray emission during the flaring state. But in the case of the quiescent state, the X-ray emission occurs at the low energy tail of the inverse Compton, which indicates CD \cite{Ajello_2020}. However, detecting LSP and ISP blazars in the hard X-ray band poses challenges due to the presence of a spectral valley between the low- and high-energy humps \citep{bottcher2007modeling}. Despite these challenges, our multi-wavelength SED modeling suggests different trends across four epochs, as depicted in Table \ref{tab:3}. Our derived bulk Lorentz factor and photon indices generally agree with previous studies using the jet-set model. However, our study indicates X-ray emission falls within the lower energy range of the inverse Compton scattering spectrum in its quiescent state (S1, S3). Alternatively, during the flaring state (S2, S4) X-ray emission shifts from inverse Compton scattering to the high-energy tail of synchrotron emission. A previous study by \cite{10.1093/mnras/stac1009,inproceedings} did not investigate X-ray emission behavior during quiescent and flaring states. Consistent with \citet{2017MNRAS.465..180L}, our independent SED modeling yields a Doppler factor of 21.5. This finding underscores the importance of understanding the bulk Lorentz factor in interpreting changes in jet power. The emission from blazars during outbursts is primarily attributed to relativistic shocks propagating through the jet. As shocks traverse the relativistic jet plasma, they boost nonthermal emission through Doppler enhancement, consistent with the theoretical framework established by \citet{1979ApJ...232...34B}. Blazars offer a unique opportunity to study relativistic jets, supermassive black holes (SMBHs), and accretion disks. To study jet power, consideration of the magnetic field, emission region size, and $\gamma_{min}$ is crucial \citep{Tantry:2024qvs}. Further observational and theoretical studies are necessary to understand the mechanisms driving these phenomena fully.

%\documentclass{article}

% Define the \aap command to expand to "Astronomy & Astrophysics"
\newcommand{\aap}{Astronomy \& Astrophysics}
\newcommand{\apj}{The Astrophysical Journal}
\newcommand{\ssr}{Space Science Reviews}
\newcommand{\mnras}{Monthly Notices of the Royal Astronomical Society}
\newcommand{\apjl}{The Astrophysical Journal Letters}
\newcommand{\pasp}{Publications of the Astronomical Society of the Pacific}
%\begin{document}

%\cite{example}

% Your bibliography and document content here

%\end{document}

%$\textunderscore$
\section{Acknowledgements}
 We sincerely thank the anonymous referees for their valuable feedback and suggestions, which have significantly improved the quality of our manuscript.The authors acknowledges the support of Prof. Ranjeev Misra in obtaining the Convolution codes used in this work . J. Tantry, Z. Shah and N. Iqbal are thankful to the Indian Space Research Organisation, Department of Space, Government of India (ISRO-RESPOND, , DS$\_$2B-13013(2)/8/2020-Sec.2) for the financial support. ZS is supported by the Department of Science and Technology, Govt. of India, under the INSPIRE Faculty grant (DST/INSPIRE/04/2020/002319). We express our gratitude to the Inter-University Centre for Astronomy and Astrophysics (IUCAA) in Pune, India, for the support and facilities provided. A. Sharma is grateful to Prof. Sakuntala Chatterjee at S.N. Bose National Centre for Basic Sciences, for providing the necessary support to conduct this research.

\bibliographystyle{elsarticle-harv} 
\bibliography{example}

%% else use the following coding to input the bibitems directly in the
%% TeX file.

%%\begin{thebibliography}{00}

%% \bibitem[Author(year)]{label}
%% For example:

%% \bibitem[Aladro et al.(2015)]{Aladro15} Aladro, R., Martín, S., Riquelme, D., et al. 2015, \aas, 579, A101
\appendix
\section{Supplementary Data}
Swift observations.

\begin{table*}
\centering
\setlength{\extrarowheight}{2pt}
\setlength{\tabcolsep}{4pt}
\begin{tabular}{c c c c c c}
\hline 
\hline
S. No. & Instrument & Observation ID & Time (MJD) & XRT exposure (ks) & UVOT exposure (ks)\\
\hline
%\hline
1. & \emph{Swift}-XRT/UVOT &00095124013&58748.03&1.14&1.14\\
2. & \emph{Swift}-XRT/UVOT&00036393021&58957.05&1.87&1.86\\
3.  & \emph{Swift}-XRT/UVOT& 00036393020&58954.01&1.98&1.98\\
4. & \emph{Swift}-XRT/UVOT & 00095124027&58804.5&1.57&1.57\\
5. & \emph{Swift}-XRT/UVOT &00095124025&58797.2&1.50&1.49\\
6. & \emph{Swift}-XRT/UVOT &00095124023&58790.07&1.22&1.22\\
7. & \emph{Swift}-XRT/UVOT &00095124028&58808.04&0.97&0.96\\
8. & \emph{Swift}-XRT/UVOT &00095124033&58829.12&1.46&1.45\\
9. & \emph{Swift}-XRT/UVOT &00095124036&58839.66&1.42&1.41\\
10. & \emph{Swift}-XRT/UVOT &00095124034&58832.31&1.38&1.38\\
11. & \emph{Swift}-XRT/UVOT &00095124038&58846.30&1.29&1.28\\
12 & \emph{Swift}-XRT/UVOT &00095124040&58853.00&1.15&1.15\\
13 & \emph{Swift}-XRT/UVOT &00095124041&58857.91&0.92&0.91\\
14. & \emph{Swift}-XRT/UVOT&00095124044&58874.74&1.11&1.10\\
15. & \emph{Swift}-XRT/UVOT &00095124037&58843.25&1.26&1.25\\
16. & \emph{Swift}-XRT/UVOT &00095124030&58818.51&1.26&1.25\\
17. & \emph{Swift}-XRT/UVOT &00095124032&58825.08&1.23&1.22\\
18. & \emph{Swift}-XRT/UVOT &00095124035&58836.02&1.20&1.19\\
19. & \emph{Swift}-XRT/UVOT &00095124031&58822.8&1.19&1.18\\
20. & \emph{Swift}-XRT/UVOT &00095124042&58860.03&1.12&1.11\\
21. & \emph{Swift}-XRT/UVOT & 00095124045  & 58878.33 & 1.22 & 1.23\\
22. & \emph{Swift}-XRT/UVOT & 00095124047 & 58885.22& 1.10 & 1.09\\
23. & \emph{Swift}-XRT/UVOT & 00095124048 & 58888.00 & 1.22 & 1.22\\
24. & \emph{Swift}-XRT/UVOT & 00095124049 &58892.00 & 1.25 & 1.25\\
25. & \emph{Swift}-XRT/UVOT & 00095124050 & 58895.03& 1.23 & 1.23\\
26. & \emph{Swift}-XRT/UVOT&00095124051&58899.61&0.10&0.09\\
27. & \emph{Swift}-XRT/UVOT&00095124052&58902.73&1.27&1.26\\
28. & \emph{Swift}-XRT/UVOT&00095124053&58906.64&1.18&1.17\\
29. & \emph{Swift}-XRT/UVOT&00095124054&58909.10&1.18&1.17\\
30. & \emph{Swift}-XRT/UVOT&00095124055&58913.08&1.13&1.12\\
31. & \emph{Swift}-XRT/UVOT&00095124056&58920.70&0.94&0.93\\
32. & \emph{Swift}-XRT/UVOT&00095124057&58923.49&1.43&1.42\\
33. & \emph{Swift}-XRT/UVOT&00095124058&58927.79&1.10&1.09\\
34. & \emph{Swift}-XRT/UVOT&00095124059&58930.31&1.37&1.36\\
35. & \emph{Swift}-XRT/UVOT&00095124060&58934.03&0.41&0.40\\
36. & \emph{Swift}-XRT/UVOT&00095124061&58934.03&1.19&1.18\\
37. & \emph{Swift}-XRT/UVOT & 00036393022 & 59667.29 & 1.19 & 1.18\\
38. & \emph{Swift}-XRT/UVOT & 00036393023 &59669.15  & 1.91 & 1.90\\
39. & \emph{Swift}-XRT/UVOT & 00036393024 & 59670.60 & 1.91 & 1.90\\
40. & \emph{Swift}-XRT/UVOT & 00036393025 &59676.03 & 2.11 & 2.12\\
41. & \emph{Swift}-XRT/UVOT & 00036393026& 60123.87 & 1.67 & 1.64\\
42. & \emph{Swift}-XRT/UVOT & 00036393027& 60127.31 & 0.33 & 0.33\\
43. & \emph{Swift}-XRT/UVOT & 00036393028& 60131.20 & 1.24 & 1.21\\

\hline
%\hline
\end{tabular}
\vspace{0.5cm}
\caption{The list of \emph{Swift}-XRT/UVOT observations used in this work is shown above. All observations are accessible}
\label{tab:4}
\end{table*}

\newpage
\begin{table*}
\centering
\setlength{\extrarowheight}{4pt}
\setlength{\tabcolsep}{4pt}
\begin{tabular}{llccccc}
\hline
\multirow{2}{*}{Source name} & \multirow{2}{*}{4FGL name} & \multirow{2}{*}{Source class} & \multirow{2}{*}{Light curve binning} & \multirow{2}{*}{QPO period } & \multirow{2}{*}{QPO cycle number} & \multirow{2}{*}{Reference} \\
[+2pt]
 & & & (days) & (days) & & \\
(1) & (2) & (3) & (4) & (5) & (6) & (7) \\
[+4pt]
\hline

B2 1520+31 & 4FGL J1522.1+3144 & FSRQ & 7 &179$\pm$42 & 6 & a\\
 & & & 7, 3 & 71$\pm$15 & 14 & a,b \\
 & & & 7 & 39$\pm$11 & 17 & a \\
PKS 1510-089 & 4FGL J1512.8-0906 & FSRQ & 0.125 & 3.6$\pm$0.07 & 5 & c \\
 & & & 7 & 92$\pm$1.2 & 7 & c\\
3C 454.3 & 4FGL J2253.9+1609 & FSRQ & 1 & 47$_{-0.51}^{+0.97}$ & 9 & d\\
3C 279 & 4FGL J1256.1-0547 & FSRQ & 7 & 39$\pm$1 & -- & e\\
 & & & 7 & 24$\pm$1 & -- & e\\
4C +01.02 & 4FGL J0108.6+0134 & FSRQ & 7 & 122$\pm$26 & 5 & a\\
 & & & 7 & 268$\pm$54 & 4 & a \\
PKS 0402-362 & 4FGL J0403.9-3605 & FSRQ & 7 & 122$\pm$42 & 5 & a\\
 & & & 7 & 221$\pm$60 & 3 & a \\
4C +21.35 & 4FGL J1224.9+2122 & FSRQ & 7 & 66$\pm$17 & 6 & a\\
PKS 1424-41 & 4FGL J1427.9-4206 & FSRQ & 7 & 57.1$\pm$4.5 & 5 & g\\
 & & & 7 & 90$\pm$22 & 5 & a\\
 & & & 10 & $\sim$355 & -- & f\\
CTA & 4FGL J2232.6+1143 & FSRQ &7 & 178$\pm$40 & 5 & a\\
 & & & 7 & 366$\pm$81 & 3 & a\\
 & & & 0.125 & 7.6$_{-0.25}^{+0.36}$ & 8 & h \\
PKS 0336-01 & 4FGL J0339.5-0146 & FSRQ & 7 & 94.6$\pm$6.8 & 6 & g \\
PKS 2255-282 & 4FGL J2258.1-2759 & FSRQ & 7 & 93$\pm$2.6 & -- & i\\
PKS 0346-27 & 4FGL J0348.5-2749 & FSRQ & 1 & 100 & -- & j\\
S5 1044+71 & 4FGL J1048.4+7143 & FSRQ & 7 & 1127$\pm$226 & 3 & a\\
 & & & 7 & 117$\pm$38 (7d LC) & 4 & a\\
4C +28.07 & 4FGL J0237.8+2848 & FSRQ & 7 & 244$\pm$88 & 3 & a\\
PKS 0454-234 & 4FGL J0457.0-2324 & FSRQ & 7 & 69$\pm$21  & 4 & a\\
3C 273 & 4FGL J1229.0+0202 & FSRQ & 7 & 177$\pm$38 & 4 & a\\
 & & & 7 & 97$\pm$25  & 3 & a\\
PMN J2345-1555 & 4FGL J2345.2-1555 & FSRQ & 7 & 191$\pm$44 & 4 & a\\
PKS 0244-470 & 4FGL J0245.9-4650 & FSRQ & 10 & 225$\pm$24 & 8 & k \\
4C +38.41 & 4FGL J1635.2+3808 & FSRQ & 10 & 110$\pm$50 & 4.2 & k\\
 & & & 10 & 60$\pm$9 & 5.4 & k\\
 & & & 10 & 19$\pm$2 & 5.1 & k\\
 & & & 10 & 35$\pm$4 & 5.3 & k\\
 & & & 10 & 104$\pm$7 & 6 & k\\
 & & & 10 & $\sim$227 & 3.7 & k\\
[+4pt]
\hline
\end{tabular}
\caption{\label{tab: Transient_QPO_1} A list of reported transient QPOs in blazars in the literature (Part 1).}
\end{table*}

% Second part of the table

\begin{table*}
\centering
\setlength{\extrarowheight}{7pt}
\setlength{\tabcolsep}{4pt}
\begin{tabular}{llccccc}
\hline
\multirow{2}{*}{Source name} & \multirow{2}{*}{4FGL name} & \multirow{2}{*}{Source class} & \multirow{2}{*}{Light curve binning} & \multirow{2}{*}{QPO period } & \multirow{2}{*}{QPO cycle number} & \multirow{2}{*}{Reference} \\
[+2pt]
 & & & (days) & (days) & & \\
(1) & (2) & (3) & (4) & (5) & (6) & (7) \\
[+4pt]
\hline

PKS 0426-380 & 4FGL J0428.6-3756 & BL Lac & 7 & 85$\pm$26 & 8 & a \\
PKS 0447-439 & 4FGL J0449.4-4350 & BL Lac & 7 & 111$\pm$42 & 7 & a\\
S5 0716+714 & 4FGL J0721.9+7120 & BL Lac & 7 & 324$\pm$77 & 5 & a\\
 & & & 5 & 31.3$\pm$1.8 & 7 & l\\
1H 1013+498 & 4FGL J1015.0+4926 & BL Lac & 7 & 52$\pm$12  & 12 & a \\
 & & & 7 & 100$\pm$25 & 4 & a\\
 & & & 7 & 264$\pm$59 & 4 & a\\
Mrk 501 & 4FGL J1653.8+3945 & BL Lac & 7 & 326$\pm$76 & 7 & a\\
PKS 0537-441 & 4FGL J0538.8-4405 & BL Lac & 7 & 55$\pm$3.3 & 7 & g \\
 & & & 7 & 54.7$\pm$3.3  & 7 & g\\
PKS 2155-304 & 4FGL J2158.8-3013 & BL Lac & 7 & 341$\pm$106 & 4 & a\\
Mrk 421 & 4FGL J1104.4+3812 & BL Lac & 7 & 300$\pm$65 & 3 & a \\
PKS 2247-131 & 4FGL J2250.0-1250 & BCU & 5 & 34.5$\pm$1.5 & 6 & m\\
 & & & 7 & 214$\pm$43 & 6 & a\\
S4 0954+658 & 4FGL J2345.2-1555 & Transition blazar & 10& 66$\pm$4.8 & 9 & n\\
 & & & 10 & 208$\pm$43 & 5 & n\\
\textbf{S5 1803+784} & \textbf{4FGL J1800.6+7828} & \textbf{Transition blazar} & \textbf{3} & \textbf{$411\pm79$} & \textbf{3} & \textbf{Our study}\\
[+4pt]
\hline
\end{tabular}
\caption{\label{tab: Transient_QPO_2} Continued from Table \ref{tab: Transient_QPO_1}. Column (1) is the name of object; Column (2) is the Fermi 4FGL name; Column (3) is the source class type; Column (4) is the transient QPO timescale and associated uncertainty in days; Column (5) is the QPO cycle number; Column (6) is the reference: (a) \cite{ren2023quasi}, (b) \cite{gupta2019detection}, (c) \cite{roy2022transient}, (d) \cite{sarkar2021multiwaveband}, (e) \cite{sandrinelli2016quasi}, (f) \cite{yang2021quasi}, (g) \cite{chen2024transient}, (h) \cite{sarkar2020multi}, (i) \cite{sharma2024detection}, (j) \cite{prince2023quasi}, (k) \cite{das2023detection}, (l) \cite{chen202231}, (m) \cite{zhou201834}, (n)\cite{gong2023two}}
\end{table*}

\end{document}